\documentclass[twocolumn,english,prl]{revtex4-1}
\usepackage[latin9]{inputenc}
\usepackage[active]{srcltx}
\usepackage{color}
\usepackage{babel}
\usepackage{amsmath}
\usepackage{amssymb}
\usepackage{graphicx}
\usepackage[unicode=true,pdfusetitle,
 bookmarks=false,
 breaklinks=true,pdfborder={0 0 0},pdfborderstyle={},backref=false,colorlinks=true]
 {hyperref}
\hypersetup{
 colorlinks,linkcolor=red,citecolor=blue}
\usepackage{breakurl}
\usepackage{lipsum}
\usepackage{dcolumn}% Align table columns on decimal point

\makeatletter

\usepackage{amsfonts}
\makeatother

\begin{document}

\title{Magnon-photon strong coupling for tunable microwave circulators}

\author{Na Zhu, Xu Han, Chang-Ling Zou$^\dagger$, Mingrui Xu, and Hong X. Tang}
\email{hong.tang@yale.edu}
\altaffiliation{$\dagger$ Current address: Department of Optics, University of Science and Technology of China, Hefei, China 230026}

\address{Department of Electrical Engineering, Yale University, New Haven, Connecticut 06520, USA}

\begin{abstract}
We present a generic theoretical framework to describe non-reciprocal microwave circulation in a multimode cavity magnonic system and assess the optimal performance of practical circulator devices. We show that high isolation ($>$ 56 dB), extremely low insertion loss ($<$ 0.05 dB), and flexible bandwidth control can be potentially realized in high-quality-factor superconducting cavity based magnonic platforms. These circulation characteristics are analyzed with materials of different spin densities. For high-spin-density materials such as yttrium iron garnet, strong coupling operation regime can be harnessed to obtain a broader circulation bandwidth. We also provide practical design principles for a highly integratible low-spin-density material (vanadium tetracyanoethylene) for narrow-band circulator operation, which could benefit noise-sensitive quantum microwave measurements. This theory can be extended to other coupled systems and provide design guidelines for achieving tunable microwave non-reciprocity for both classical and quantum applications.

\end{abstract}

\maketitle
\section{Introduction}
Non-reciprocal microwave devices are ubiquitous and important in the classical and quantum information processing, as they protect delicate measurements from reflected signals \cite{caloz2018electromagnetic}. The non-reciprocal effect arises from broken time-reversal symmetry, traditionally realized with ferrite materials \cite{linkhart2014microwave,fay1965operation,wu1974wide}. Recently, a variety of avenues have been reported to realize non-reciprocity without the use of magnetic materials, including optomechanical coupling \cite{bernier2017nonreciprocal,peterson2017demonstration}, reservoir engineering \cite{fang2017generalized,metelmann2015nonreciprocal}, nonlinear effect \cite{chapman2017widely,chapman2019design}, and temporal modulation \cite{sounas2017non}. Those approaches, albeit being non-magnetic, typically require strict phase matching condition and have limited tunability \cite{fang2017generalized,chapman2019design,chapman2017widely,peterson2017demonstration,bernier2017nonreciprocal}, with added complexity in experimental implementations. Nowadays, due to the high demand in sensitive microwave signal detections, especially at single-photon level for superconducting quantum circuits, low-loss, tunable, and compact electromagnetic circulator devices are of great interest \cite{caloz2018electromagnetic}.

Cavity magnonic systems have attracted significant attentions recently \cite{stancil2009spin,tabuchi2015coherent,zhang2015magnon,zhang2016cavity,zhang2016cavity,liu2016optomagnonics,osada2016cavity,kostylev2016superstrong,bhoi2014study,bhoi2017robust,match2019transient,harder2017topological,zhang2014electric} due to the strong interaction between magnon excitations and microwave photons. Previous studies have demonstrated magnon-photon strong coupling in various resonant microwave systems, such as copper 3-dimentional (3D) cavities \cite{zhang2015magnon,zhang2016cavity,zhang2016cavity,liu2016optomagnonics,osada2016cavity,kostylev2016superstrong} and coplanar microwave circuits \cite{bhoi2014study,bhoi2017robust,li2019strong}. However, those cavity magnonic systems are in the conventional coherent coupling configuration, where magnons are coupled with a single microwave mode without any non-reciprocal effect. Only a few recent works have investigated non-reciprocal coherent or dissipative magnon-photon coupling in two-port systems to demonstrate isolator devices \cite{wang2019nonreciprocity,zhang2019strong}. The study of the three-port non-reciprocal magnonic platform is highly motivated because of its applications in sensitive cryogenic microwave reflection measurements \cite{corcoles2015demonstration,kono2018quantum}. 

In this work, we present a generic theoretical model for non-reciprocal multimode cavity magnonic systems. We show that by harnessing the selective coupling between the magnon mode and microwave modes with different chiralities, as well as the interference effect between different paths in a three-port system, non-reciprocal microwave circulation with high isolation, low insertion loss, and flexible controllability can be achieved. For device implementation, we propose a practical design based on a high-quality factor (\textit{Q}) superconducting ring resonator which is coupled with a high-\textit{Q} magnon mode in a low-Gilbert-damping magnetic media \cite{stancil2009spin, sparks1964ferromagnetic} under a bias magnetic field. Two exemplary material platforms are discussed: (1) yttrium iron garnet (YIG), a high-spin-density material that can work in the strong coupling regime to obtain broader circulation bandwidth \cite{zhang2016x}, and (2) vanadium tetracyanoethylene ($\textrm{V[TCNE]}_2$), a highly integratible low-spin-density material for narrow-bandwidth operation \cite{yu2014ultra}. Unlike commercial circulators designed for octave broadband operations, this work exploits cavity enhanced circulation effect and trades the circulation bandwidth for high isolation and low insertion loss, which are the most desirable performance parameters for delicate single-photon level quantum measurements. 

\section{Theoretical Model}
\subsection{Coupled-Mode Theory}
\begin {figure}[htbp]
\centering
\includegraphics[width=\linewidth]{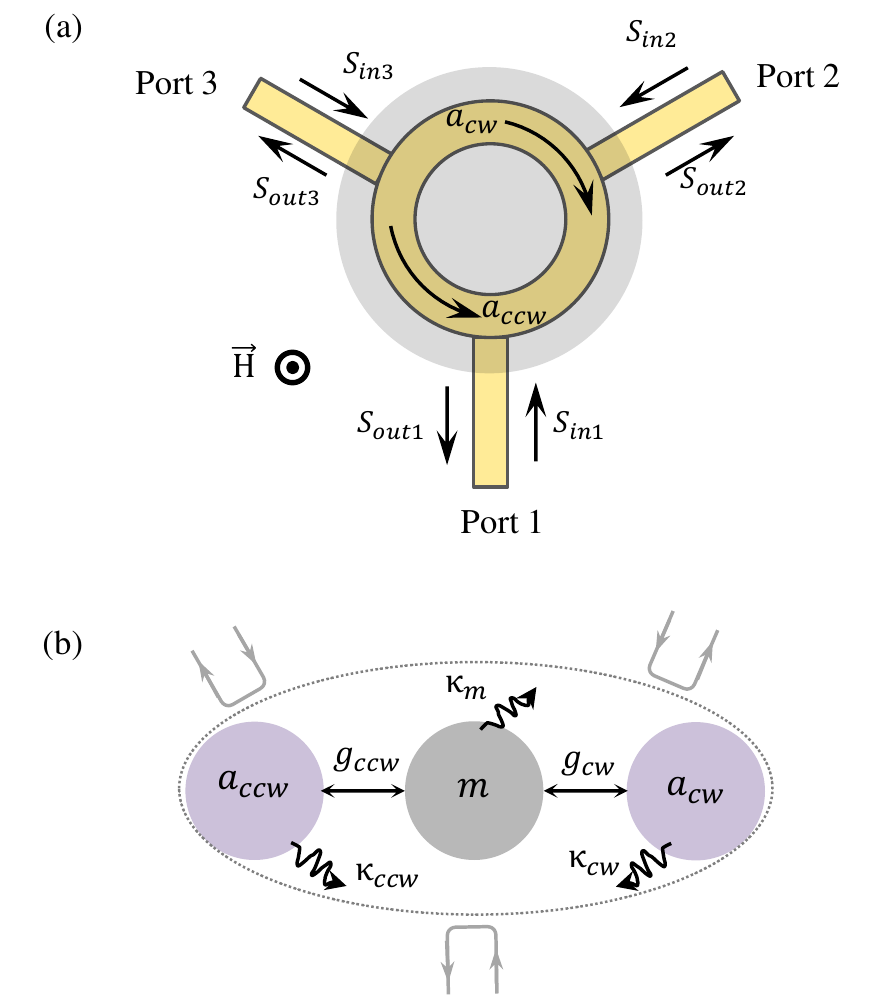}
\caption{ (a) Schematic diagram of the device. The
superconducting ring resonator supports both counter-clockwise (CCW, ${a_{ccw}}$) and clockwise (CW, ${{a}_{cw}}$) rotating microwave modes. A ferrimagnetic disk (grey color) with similar dimension is placed on top of the superconducting ring and biased perpendicularly. There are three straight waveguides with three-fold geometrical symmetry inductively coupled with the ring resonator. (b) The schematic shows the magnon-photon coupling, where the microwave mode $a_{ccw}$ is coupled with the magnon mode ${m}$ with the coupling strength $g_{ccw}$, while mode $a_{cw}$ is coupled with magnon mode via a different coupling strength $g_{cw}$. $\kappa_{ccw}$, $\kappa_{cw}$, and $\kappa_{m}$ are the total dissipation rates for mode $a_{cw}$, $a_{ccw}$, and $m$, respectively.}
\label{fig1}
\end {figure}

A schematic of the circulator is shown in Fig. \ref{fig1}(a), where a three-port superconducting ring resonator simultaneously supports two degenerate counter-rotating microwave modes. This ring resonator is aligned with a ferrimagnetic disk of similar dimension for optimal mode overlap. Under a static out-of-plane magnetic bias field, the ferrimagnetic disk supports a uniform magnon mode with the resonant frequency linearly proportional to the external field \cite{walker1957magnetostatic}, $\omega_\mathrm{m}\approx\gamma\left|\vec{B_\mathrm{o}}\right|$, where $\gamma = 2.8 {\text{ MHz/Oe}}$ is the gyromagnetic ratio. The system Hamiltonian can be written as 
\begin{equation}
H/\hbar=\omega_{ccw}{a_{ccw}}^{\dagger}{a_{ccw}}+\omega_{cw}{a_{cw}}^{\dagger}a_{cw}+\omega_{m}m^{\dagger}m+H_{int}/\hbar.
\label{eq:refname1}
\end{equation}
Here, ${a_{ccw}}\left({a_{ccw}}^{\dagger}\right), {a_{cw}}\left({a_{cw}}^{\dagger}\right), m\left(m^{\dagger}\right)$ are the annihilation (creation) operators for the counter clockwise (CCW) and the clockwise (CW) microwave mode, and the magnon mode, respectively, with their resonant frequencies denoted as $\omega_{ccw}$, $\omega_{cw}$, and $\omega_{m}$. Since the CCW and the CW modes are orthogonal, we only need to consider their linear coupling with the magnon mode in our system. So the interaction Hamiltonian is 
\begin{equation}
\begin{split}
H_{int}/\hbar=&-g_\mathrm{ccw}\left({a_{ccw}}+{a_{ccw}}^{\dagger}\right)\left(m+m^{\dagger}\right)\\&-g_\mathrm{cw}\left({a_{cw}}+{a_{cw}}^{\dagger}\right)\left(m+m^{\dagger}\right),
\end{split}
\label{eq:refname2}
\end{equation}
where $g_\mathrm{ccw}$ and $g_\mathrm{cw}$ are the coupling strengths between the respective microwave mode and magnon mode \cite{zhang2019strong,li2018magnon,zhang2014strongly}.

Under the rotating wave approximation (RWA), the Heisenberg-Langevin equation can be written as
\begin{equation}
\dot{\boldsymbol{a}}=\boldsymbol{{{M}_{1}}a}+\boldsymbol{K}\boldsymbol{s}_{in},
\label{eq:refname3}
\end{equation}
with the input-output relation
\begin{equation}
\boldsymbol{s}_{out}=\boldsymbol{C}\boldsymbol{s}_{in}+\boldsymbol{{M}_{2}}\boldsymbol{a}.
\label{eq:refname4}
\end{equation}
Here $\boldsymbol{a}=\left\{ a_{ccw},a_{cw},m\right\} ^{T}$ is the vector of the cavity field. $\boldsymbol{s}_{in}=\left\{ s_{in1},s_{in2},s_{in3}\right\} ^{T}$ and $\boldsymbol{s}_{out}=\left\{ s_{out1},s_{out2},s_{out3}\right\} ^{T}$ are the input and the output fields at the three ports. 
Matrix $\boldsymbol{{M}_{1}}(3\times3)$ is given as
\begin{equation}
\boldsymbol{{M}_{1}}=\left(\begin{array}{ccccc}
-i\omega_{ccw}-\frac{\kappa_{ccw}}{2} & 0 &ig_{ccw} \\ 0 & -i\omega_{cw}-\frac{\kappa_{cw}}{2} & ig_{cw}\\
ig_{ccw}&ig_{cw}&-i\omega_{m}-\frac{\kappa_{m}}{2}
\end{array}\right),
\label{eq5}
\end{equation}
in which $\kappa_{{ccw}}$, $\kappa_{{cw}}$, and $\kappa_{{m}}$ are the total dissipation rates for the microwave and magnon modes, respectively. $\boldsymbol{K}$ and $\boldsymbol{{{M}_{2}}}$ are the $(3\times3)$ matrices describing coupling of three incoming/outgoing waves with two resonant modes. Based on relation among $\boldsymbol{K}$, $\boldsymbol{C}$, and $\boldsymbol{{{M}_{2}}}$ (see Appendix), the matrix $\boldsymbol{K}(3\times3)$ satisfies the energy conservation relation $\boldsymbol{K}\boldsymbol{K}^{\dagger}=\boldsymbol{\Gamma_{e}}$, and can be in general written as $\boldsymbol{K}=-\boldsymbol{{M_{2}}^{\dagger}}\boldsymbol{C}=$
\begin{equation}
\begin{split}
\left(\begin{array}{ccccc}
\sqrt{\kappa_{ccw,e1}}&\sqrt{\kappa_{ccw,e2}}e^{i\alpha}&\sqrt{\kappa_{ccw,e3}}e^{i\eta}\\\sqrt{\kappa_{cw,e1}}e^{i\beta_{1}}&\sqrt{\kappa_{cw,e2}}e^{i\beta_{2}}&\sqrt{\kappa_{cw,e3}}e^{i\beta_{3}}\\0&0&0
\end{array}\right).
\end{split}
\label{eq:refname6}
\end{equation}
Here $\alpha$($\eta$) is the relative phase between the excitation port 2(3) and port 1  for mode $a_{ccw}$. $\beta_{1}$ describes the phase difference between CW and CCW modes at port 1, and ($\beta_{2(3)}$ $-$ $\beta_{1}$) denotes the relative phase difference for the CW mode between port 2(3) and 1. $\kappa_{ccw,e1(2,3)}$ and $\kappa_{cw,e1(2,3)}$ represent the external coupling rates of the two microwave modes to the three input/output ports, respectively.

$\boldsymbol{C}$ is the $(3\times3)$ matrix describing direct coupling of incoming and outgoing waves. Due to energy conservation, $\boldsymbol{C}$ must be unitary $\boldsymbol{C}^{\dagger}\boldsymbol{C=I}$. But the specific expression for $\boldsymbol{C}$ depends on the physical implementation of the excitation ports. For example, in the case of a waveguide end-coupling scheme with negligible crosstalk between ports, $\boldsymbol{C=I}$ for open-ended (capacitive) coupling \cite{pozar2009microwave,adamyan2016tunable}; and $\boldsymbol{C=-I}$ for short-ended (inductive) coupling \cite{cernrf,pozar2009microwave,bothner2013inductively,sierra1999study}. In this paper, we assume that the waveguides and the ring are inductively coupled with no dissipative coupling across different ports, namely,
\begin{equation}
\boldsymbol{C}=\left(\begin{array}{ccccc}
-1 & 0 &0 \\ 0 & -1&0\\0 & 0&-1
\end{array}\right).
\end{equation}
With the expressions of matrices in Eq. \ref{eq:refname3} \& \ref{eq:refname4}, the equation of motion can be solved in frequency domain to obtain the scattering matrix (defined by $\boldsymbol{s}_{out}[\omega]=\boldsymbol{S}[\omega]\boldsymbol{s}_{in}[\omega]$) 
\begin{equation}
\boldsymbol{S}[\omega]=\boldsymbol{C}+\boldsymbol{M_2}\left[-i\omega \boldsymbol{I}-\boldsymbol{M_1}\right]^{-1}\boldsymbol{K}.
\label{eq:refname8}
\end{equation}

\subsection{Circulation Under Three-fold Symmetry}
Based on the generic coupled mode theory described above, we now focus our discussion on the system circulation with three-fold geometrical rotational symmetry. Under this condition, the microwave modes $a_{ccw}$ and $a_{cw}$ are degenerate and have external coupling rates to each of the three waveguides with the value ${{\kappa}_{e}}/{3}$. The total dissipation rate $\kappa_{{ccw}}$ $=$ ${\kappa_{{cw}}=\kappa_{{e}}+\kappa_{{i}}}$, where ${\kappa_{\mathrm{i}}}$ and $\kappa_{\mathrm{e}}$ are the microwave intrinsic dissipation and total external coupling rates, respectively. 

At the same time, we have the relative excitation phase difference between port 2(3) and 1 as $\alpha$ = $-\eta$ = $-{2n\pi}/{3}$, where $n$ is the integer represents the mode number. Here, we focus on the two degenerate fundamental circulation microwave modes ($n=1$), for which we have $\omega_{ccw}$ $=$ $\omega_{cw}$ and $\alpha$ = $-\eta$ = $-{2\pi}/{3}$. Such phase difference is determined by the three-fold rotational symmetry, because the fundamental mode is formed when the wavelength equals ring perimeter with azimuthal number to be 1, leading to 2$\pi$ phase shift along the ring \cite{linkhart2014microwave,wu1974wide}. The relative pahse difference between two ports under three-fold rotational symmetry will be $\pm{2\pi}/{3}$, depending on the mode propagating directions. Since these two modes are orthogonal to each other, it can be shown that the relative excitation phase $\beta_{1}$ between $a_{ccw}$ and $a_{cw}$ does not contribute to the final expression of the scattering matrix, and we can set that to be 0 for simplicity.

%Hence, for the degenerate fundamental CW mode, the relative phase differences between port 2(3) and 1 are $\pm2\pi/3$, respectively.

%The CCW and CW resonant frequencies are mostly determined by the geometrical dimension of the ring, and are highly degenerate unless extra geometrical perturbations are introduced. The coupling rates with each port ${{\kappa}_{e1(2,3)}}$ are dominated by the impedance of the waveguides and ring. By utilizing three-fold symmetric superconducting circuit patterning, the degenerate resonant frequencies and same external port coupling rates can be achieved experimentally without noticeable differences \cite{wolff1972microstrip,wu2008design,han2002strong}. For the microwave intrinsic dissipation rate $\kappa_{{i}}$ of the superconducting ring, this value is minimal at the cryogenic temperature \cite{xu2019frequency}, and can be regarded as evenly dissipated to three ports considering of the system symmetry.

Due to the selective coupling rule, the magnon mode would only couple with microwave mode with the same chirality. Therefore, for our two circular microwave modes, only $g_{ccw}=g$ is significant and $g_{cw}\approx0$. It's worth pointing out that a different eigenmode basis can be chosen for the two degenerate microwave modes by applying a unitary rotation $\boldsymbol{U}(\theta)=\left(\begin{array}{ccc}
\textrm{cos}\theta &\textrm{-sin}\theta&0\\
\textrm{sin}\theta & \textrm{cos}\theta&0\\0&0&1
\end{array}\right)$. Then the new $\boldsymbol{M_1}$ matrix will become the general form in Eq.\,\ref{eq5} with $g_{ccw}=g\sin{\theta}$ and $g_{cw}=-g\cos{\theta}$. This rotation of the basis, nevertheless, won't change the physical results of our analysis.

Under the three-fold geometrical symmetry, the coupling matrix $\boldsymbol{K}$ is given by 
\begin{equation}
\boldsymbol{K}=\sqrt{\frac{\kappa_{e}}{3}} \left(\begin{array}{ccccc}
1&e^{-i\frac{2\pi}{3}}&e^{i\frac{2\pi}{3}}\\1&e^{i\frac{2\pi}{3}}&e^{-i\frac{2\pi}{3}}\\0&0&0
\end{array}\right)
\label{eq:refname9}
\end{equation}
Finally, using Eq. \ref{eq:refname8}, we can obtain the scattering matrix elements

\begin{equation} \label{eq14}
\begin{split}
{S}_{11} = \\&S_{22} = S_{33} = -1+\frac{{\kappa}_{e}}{3({{\kappa}_{cw}}/{2}-i(\omega-{\omega}_{cw}))}\\
&+\frac{{\kappa}_{e}({\kappa}_{m}/2-i(\omega-{\omega}_{m}))}{{3({g}}^{2}+({\kappa}_{m}/2-i(\omega-{\omega}_{m}))({\kappa}_{ccw}/2-i(\omega-{\omega}_{ccw})))}
\end{split}
\end{equation}

\begin{equation} \label{eq15}
\begin{split}
{S}_{12} = \\&S_{23} = S_{31} = \frac{{{e}^{\frac{i2\pi}{3}}\kappa}_{e}}{3({{\kappa}_{cw}}/{2}-i(\omega-{\omega}_{cw}))}\\&+\frac{{e}^{\frac{-i2\pi}{3}}{\kappa}_{e}({\kappa}_{m}/2-i(\omega-{\omega}_{m}))}{{3({g}}^{2}+({\kappa}_{m}/2-i(\omega-{\omega}_{m}))({\kappa_{ccw}}/2-i(\omega-{\omega}_{ccw})))}
\end{split}
\end{equation}

\begin{equation} \label{eq16}
\begin{split}
{S}_{21} = \\&S_{13} = S_{32} = \frac{{{e}^{\frac{-i2\pi}{3}}\kappa}_{e}}{3({{\kappa}_{cw}}/{2}-i(\omega-{\omega}_{cw}))}\\&+\frac{{e}^{\frac{i2\pi}{3}}{\kappa}_{e}({\kappa}_{m}/2-i(\omega-{\omega}_{m}))}{{3({g}}^{2}+({\kappa}_{m}/2-i(\omega-{\omega}_{m}))({\kappa}_{ccw}/2-i(\omega-{\omega}_{ccw})))}
\end{split}
\end{equation}

We see that the scattering matrix satisfies relations $S_{11}$ = $S_{22}$ = $S_{33}$, $S_{12}$ = $S_{23}$ = $S_{31}$, and $S_{21}$ = $S_{13}$ = $S_{32}$, as expected from the rotational symmetry. Without the three-fold symmetry, for example, if three waveguides are design to have different dimensions, the microwave external coupling rates for three ports are very different, the reflection coefficients could be in general different $S_{11}$ $\ne$ ${S}_{22}$ $\ne$ ${S}_{33}$. Such degeneracy will also be broken for other transmission parameters. Hence, if the microwave circuit is designed to be non-symmetric, the system circulation performance is only optimized for one kind of connection configuration under certain magnetic bias condition.

Next, as we can see from Eq. (\ref{eq14}, \ref{eq15}, \ref{eq16}), although the microwave CCW and CW mode are degenerate due to the structural symmetry, selective coupling between the magnon mode and the two rotational microwave modes gives rise to a non-reciprocal scattering matrix. If there is no magnon-photon coupling ($g$ $=$ 0), $S_{12}$ and $S_{21}$ do not have the amplitude non-reciprocity. In the situation when ($g$ $\ne$ 0), the signal circulation starts to occur due to magon-coupling-induced interference, the isolation ratio between the scattering matrix parameters $S_{12}$ and $S_{21}$ depends on both $g$ and the magnon resonance detuning ($\omega_{ccw/cw}$ $-$ $\omega_{m}$). 

%%if the microwave CCW and CW modes are non-degenerate in the frequency domain, the system has the non-reciprocal signal transmission without the coupling to the magnon mode. This is because once the signal propagation direction is reversed, for the same propagation path, the CCW and CW modes effectively swap, leading to different phase interference. The origin of this non-reciprocity is the different phase shift between mode $a_{ccw}$ and $a_{cw}$ at the output ports. Thus,

Lastly, we discuss the experimental controllability for the parameters to realize the three-fold symmetry. First, regarding the frequency degeneracy of CCW and CW modes, for the planar microwave ring resonators, such as the microstrip ring, the resonant frequencies for CCW and CW modes with same mode number are highly degenerate, unless the extra geometrical perturbations are introduced \cite{wu2008design,han2002strong}. Therefore, the $\omega_{ccw/cw}$ can be regarded as degenerate when the microwave circuit has three-fold geometrical symmetry. Next, for the external coupling rates among the CCW/CW mode and three waveguides ($\kappa_{ccw,e1(2,3)}$ and $\kappa_{cw,e1(2,3)}$), as the CCW and CW modes are highly degenerate, the external coupling rate between each waveguide and the ring is determined by the impedance \cite{krawczyk2011microstrip,besedin2018quality,chang2004microwave}, which depends on the geometrical dimensions and dielectric substrate material \cite{ pozar2009microwave, bahl1977designer}. Thus, by utilizing precise lithographical patterning, we can engineer the waveguides on the same chip with identical geometry yielding three-fold symmetry. The degeneracy of external coupling rates can be well controlled and modeled at the microwave frequencies. Other literatures also explored tuning external coupling rates dynamically during the measurements by utilizing SQUID \cite{peropadre2013tunable}. For the microwave intrinsic dissipation rate $\kappa_{{i}}$ of the superconducting ring, this value is minimal at the cryogenic temperature to be around several megahertz \cite{xu2019frequency}, and can be regarded as evenly dissipated to three ports considering of the system symmetry.

\section{Material Platforms}

In this session, we are interested in implementing circulators via both high spin density and low-spin density ferrimagnetic materials. To achieve circulators with low insertion loss and high isolation, an ideal ferrimagnetic material should have the low Gilbert damping factor $\zeta$. A wide exploited high density ferrimagnetic materials is  single crystal ferrimagnetic material yttrium iron garnet (YIG, $\zeta_\textrm{YIG} \sim$ $3\times10^{-5}$) \cite{sun2012growth,zhu2017patterned,chang2014nanometer,onbasli2014pulsed,dionne2009magnetic,sparks1964ferromagnetic} whereas a particularly interesting low density ferrimagnetic material is organic-based ferrimagnet vanadium tetracyanoethylene thin films ($\textrm{V[TCNE]}_{2}$, ${\zeta}_{\textrm{V[TCNE]}_{2}} \sim$ $3.8\times10^{-5}$) \cite{zhu2016low,franson2019low,yu2014ultra}. The relevant material parameters of bulk and thin-film YIG as well as V[TCNE]$_2$  are listed in table I \cite{zhang2019strong,sun2012growth,zhu2016low}.

\label{table1}
\begin{table}[b]
\begin{ruledtabular}
\begin{tabular}{lcdr}
\textrm{Name}&
{$4{\pi}M_{s}$}&
\multicolumn{1}{c}{{$\Delta{f}$ $@$ $\sim$ 10 GHz}}&
\textrm{Thickness}\\
\colrule
YIG bulk & 1750 G  & 3 \textrm{ MHz} & {$\sim$}1 mm\\
YIG thin film & 1750 G & 3 \textrm{ MHz} & 1-5 $\mu$m\\
${\textrm{V[TCNE]}}_{2}$ & 90 G & 2 \textrm{ MHz} & 1-5 $\mu$m\\
\end{tabular}
\end{ruledtabular}
\caption{Material parameters for YIG and  $\textrm{V[TCNE]}_{2}$}
\end{table}

\subsection{YIG-based Circulator}

Yttrium iron garnet (YIG) is widely used as the magnon media for its excellent magnetic properties such as long  spin  lifetime and  wide tunability. YIG has a relatively large saturation magnetization ($4{\pi}{\textrm{M}_{s}}$, 1750 G \cite{wu2010nonlinear}), which is about 20 times higher than that of $\textrm{V[TCNE]}_{2}$, thus can be regarded as a high-spin-density material \cite{sun2012growth,zhu2017patterned,chang2014nanometer,onbasli2014pulsed,dionne2009magnetic,sparks1964ferromagnetic,zhu2016low,franson2019low,yu2014ultra}. 
Recently, YIG thin films and spheres have been used to study the coherent coupling between magnons and microwave photons. The reported coupling strengths range from several megahertz to a few gigahertz by engineering the mode overlap factor, mode volume and resonant frequencies \cite{zhang2015magnon,zhang2016cavity,zhang2016cavity,liu2016optomagnonics,osada2016cavity,kostylev2016superstrong,bhoi2014study,bhoi2017robust,li2019strong}. Both YIG thin film and YIG bulk can be promising candidates for building high performance circulators, due to their low Gilbert damping factors. The key difference is that, for YIG thin film, when magnetized in the out-of-plane direction, the bias field needs to overcome the demagnetiztion field ($\sim$1750 Oe) to fully saturate and effectively excite the magnon resonances ($\omega_\mathrm{m}=\gamma\left|{\vec{B_\mathrm{o}}-\vec{H_\mathrm{d}}}\right|$, where $\vec{H_\mathrm{d}}$ is the demagnetizing field). On the other hand, the demagnetization field for the bulk is only around tens of oersted. Thus, compared with bulk, using YIG thin-film as the magnon media may introduce extra flux in the superconducting microwave resonator \cite{brandt1998superconductor,de2018superconductivity}.

\begin {figure}[htbp]
\centering
\includegraphics[width=\linewidth]{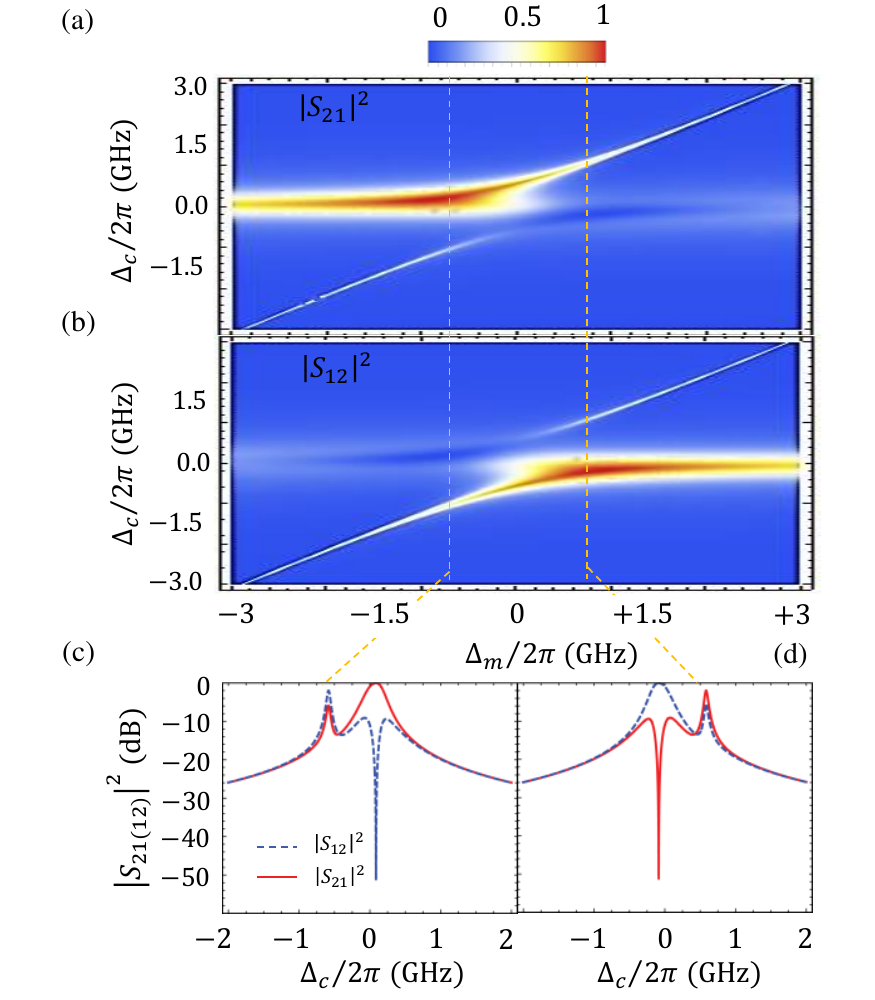}
\caption{(a) \& (b) are the mapping of the transmission ${{|S_{21}|}^{2}}$ and ${{|S_{12}|}^{2}}$ respectively as a function of ${\Delta}_{m}$ and ${\Delta}_{c}$, under strongly coupled conditions with ${\kappa}_{e}/{2\pi}$ = 600 MHz, $2{g}/{2\pi}$ = 1,200 MHz. The mode splitting at the on-resonance condition is around 2$g$. Under different detunings ${{\Delta}_{m}}/{2\pi}={\pm}0.87$ GHz, the transmission spectra are plotted in (c)/(d) when YIG is biased with magnetic resonance lower/higher than the microwave resonance.}
\label{fig2}
\end {figure}

In this session, we show the use of YIG bulk disk as the magnon media to achieve microwave circulation under the condition where magnon and photon are strongly coupled ($g$ $>$ $\kappa_{ccw}$, $\kappa_{cw}$, $\kappa_{m}$), and the superconducting microwave ring resonator is over-coupled ($\kappa_{e}$ $>$ $\kappa_{i}$). In this scenario, the magnon resonance can be detuned from the microwave resonance into the dispersive coupling regime to reduce loss induced by ferrimagnetic damping. The inherent tunability of the magnon resonance offers extra functionalities for this system to achieve adjustable isolation ratio and switchable signal propagation directions.

To be compatible with the relatively strong bias magnetic field, superconductors with high critical field, such as the NbTi film on the sapphire substrate, can be used  for the microwave circuits fabrication \cite{shapira1965upper}. The proposed device should be designed to maintain the three-fold structural symmetry with three identical waveguides inductively coupled with ring resonator at the $2\pi/3$ angle difference. The resonant frequencies of CCW and CW mode are determined by the superconducting resonator geometry, and are degenerate ($\omega_{ccw}=\omega_{cw}$) in absence of the magnon media. The intrinsic microwave loss $\kappa_{i}/2\pi$ for both microwave modes can be estimated based on the previous literature to be around several megahertz \cite{xu2019frequency,barends2007niobium, ebrahimi2016superconducting}, when the static magnetic field is much lower than the NbTi critical field. The total external coupling rate for the ring resonator can be adjusted by changing the impedance of the input waveguides. Here, based on previous magnon-photon coupling studies within microwave coplanar resonators \cite{wang2019nonreciprocity, bhoi2014study}, the external coupling rate ${\kappa_{e}}/2\pi$ can be engineered from tens of megahertz to several gigahertz. To explore both magnon-photon strong and weak coupling regimes, we will set $\kappa_{e}/2\pi$ to a moderate value (600 MHz), which can be achieved experimentally via the impedance design. Given the system's structural symmetry, the external coupling rates should be identical for both CCW and CW modes.

The magnon-photon coupling strength $g$ can be engineered by changing the microwave mode volume and frequency \cite{ li2018magnon,huebl2013high}, engineering the field overlap \cite{ li2018magnon,huebl2013high, zhang2014strongly},  and utilizing  materials with different spin-densities \cite{ li2018magnon}. During experimental measurements, $g$ can be tuned dynamically by changing the gap between the ferrimagnetic disk and the circuit to effectively tune the field overlap factor \cite{ zhang2014strongly},  adding additional ground plate to modify the microwave field distribution, and tuning the temperature \cite{ maier2017tunable}. Such engineering flexibility offers various techniques for system optimizations.

Figure \ref{fig2}(a) and (b) show a mapping of the transmission spectra  ${{|S_{21}|}^{2}}$ and ${{|S_{12}|}^{2}}$ respectively as a function of frequency detunings ${\Delta}_{m}$ ($\omega_{m}-\omega_{ccw/cw}$) and ${\Delta}_{c}$ ($\omega-\omega_{ccw/cw}$) calculated from Eq. \ref{eq15} \& \ref{eq16}, with ${g}/{2\pi}$ = 600 MHz, ${\kappa}_{e}/{2\pi}$ = 600 MHz, ${\kappa}_{i}/{2\pi}$ = 2 MHz, and ${\kappa}_{m}/{2\pi}$ = 3 MHz.  The avoided crossings in the transmission spectra indicate the strong coupling between the microwave photon and the magnon, with clear asymmetric transmission under positive and negative magnon resonance detuning  ${\Delta}_{m}$.  As we can see in the Fig. \ref{fig2} (c) \& (d), $|S_{21}(-|\Delta_{m}|)|^{2}$ equals $|S_{12}(+|\Delta_{m}|)|^{2}$. At the same time, the optimal circulation with minimal insertion loss and large isolation ratio ($|{\textrm{iso.}}|$ $=$ $20{\textrm{log}_{10}}|S_{21}|/|S_{12}|$) can be achieved by optimizing the magnon resonance detuning  ${\Delta}_{m}$. As shown in Fig. \ref{fig2} (c) \& (d), when $\Delta_{m}=\pm$ 0.87 GHz, the insertion loss ($|\textrm{IL}|$) can be as low as 0.05 dB, with the isolation ratio reaching 56 dB. Noticeably, the directionality of the signal propagation can be easily switched by changing ${\Delta}_{m}$ without reorienting the external magnetic field. When the magnon resonance is optimized, the maximal isolation ratio can be achieved in a system is limited by intrinsic losses $\kappa_{i}$ and $\kappa_{m}$. If the system dissipation losses can be optimized to ${\kappa}_{i}/{2\pi}$ $=$ ${\kappa}_{m}/{2\pi}$ $=$ 0.5 MHz by the fabrication process optimization and/or operating at ultra-low temperatures, the isolation ratio can be further enhanced to 63 dB, with insertion loss being reduced to 0.02 dB.

\begin {figure}[htbp]
\centering
\includegraphics[width=\linewidth]{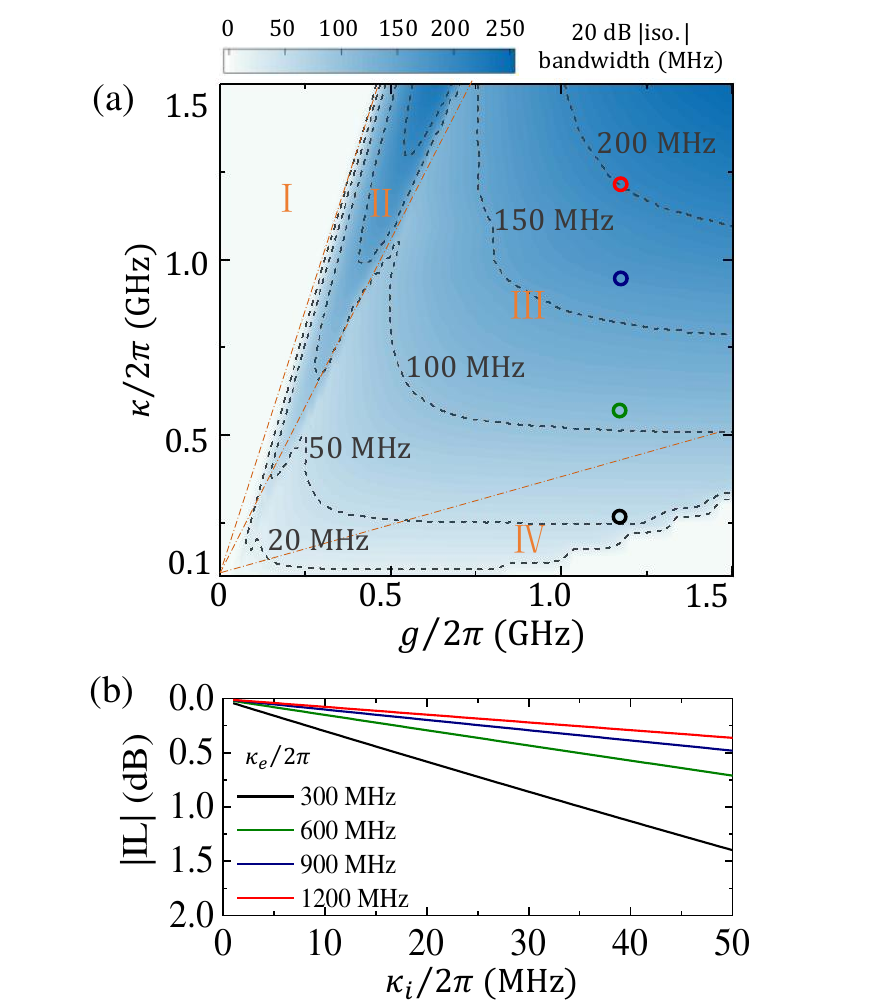}
\caption{ The 20 dB isolation ratio ($|\textrm{iso.}|$) bandwidth is plotted in (a)  as a function of ${g}/{2\pi}$ and ${\kappa}/{2\pi}$, where the black dash lines are contour lines of 20, 50, 100, and 200 MHz $|\textrm{iso.}|$ bandwidth, respectively. The magnon frequency detuning ${\Delta}_{m}$ is chosen at each $\kappa$ and $g$ to maximize the isolation bandwidth. (b) When  ${g}/{2\pi}$ is set to be 1,200 MHz, the insertion loss (IL) as a function of ${\kappa_{i}}/{2\pi}$, for different ${\kappa_{e}}/{2\pi}$ is plotted. The colored lines in (b) corresponding to the colored circles in (a).}
\label{fig3}
\end {figure}

Next, we study the controllability of the coupled system based on other tuning parameters. Figure \ref{fig3}(a) shows a mapping of 20 dB isolation bandwidth as a function of the magnon-photon coupling strength $g$, and the microwave total dissipation rate $\kappa={\kappa}_{e}+{\kappa}_{i}$. The degenerate microwave resonant frequency is 10 GHz for this calculation. The contour lines delineate 20 MHz, 50 MHz, 100 MHz, and 200 MHz circulation bandwidth when $|\textrm{iso.}|=20$ dB, respectively. The parameter spaces can be divided into four regimes. Region I represents weak coupling regime ($g \ll \kappa$) where the coupling strength is too small to induce significant phase modulation at output ports. Similarly, in region IV, at the ultra strong coupling limit when $g$ $\gg$ $\kappa$, the weak signal circulation can be understood as the CCW and CW microwave mode splitting in the frequency domain is much larger than its linewidth, due to the strong coupling with the magnon mode. Thus, the CCW and CW modes have minimal effective overlap in the frequency domain to enable the effective circulation. In region III, when the system is near/at the strong coupling regime $g$ $\sim$ $\kappa$, the 20 dB isolation bandwidth increases linearly with $g$ and $\kappa$. The optimal circulation happens when the magnon mode is detuned away from microwave modes. Lastly, in region II, when magnon and photon are weakly coupled $g<$ $\kappa$, the isolation bandwidth also yields a linear relation as $g$ and $\kappa$ are increasing linearly. Compared with region III when the system is near strongly coupled, in region II, the magnon mode is tuned close to the microwave mode to achieve strong enough phase modulation for broad band circulation. These properties show that for different applications, various isolation bandwidths can be engineered by tuning magnon resonant frequency, magnon-photon coupling strength, and microwave total dissipation rate.

Another critical performance parameter of a circulator is the insertion loss ($|\textrm{IL}|$). In Fig. \ref{fig3}(b), we study the contribution of the microwave intrinsic loss to the insertion loss at different microwave external coupling rates with the magnon-photon coupling rate fixed at $g/2\pi=\textrm{1,200 MHz}$. As ${\kappa_{e}}/2{\pi}$ is varied from 300 to 1,200 MHz, with the increase of $\kappa_{i}$, the $|\textrm{IL}|$ increases correspondingly. Thus, the high $\textit{Q}$ superconducting microwave resonator is ideal for achieving ultra low loss circulation. Fig. \ref{fig3}(b) also indicates that at same microwave intrinsic loss rate $\kappa_{i}$, the insertion loss can be reduced by increasing the microwave external coupling rate $\kappa_{e}$. This is because that when $\kappa_{e}$ is dominating in the total microwave dissipation rate, less microwave signal in the resonator dissipates into the intrinsic loss channel, thus, resulting into the low insertion loss. The analyses above establish that many desirable features of a circulator -- high isolation, low insertion loss and tunable bandwidth --- can be achieved simultaneously.

\subsection{${\textrm{V[TCNE]}_{2}}$-based Circulator}

In this session, we focus on the microwave circulation based on the low-spin density ferrimagnetic material ${\textrm{V[TCNE]}_{2}}$, which is an organic-based high quality ferrimagnetic semiconductor ($E_{g}=0.5$ eV, $\sigma=0.01$ S/m) exhibiting room temperature magnetic ordering ($T_{c}$ $>$ 600 K) \cite{,manriquez1991room,pokhodnya2000thin,yu2014ultra}. ${\textrm{V[TCNE]}_{2}}$ has very low Gilbert damping factor on the similar level as single crystal YIG for both continuous and micro-patterned films \cite{franson2019low}. Particularly, this material can be integrated onto various substrates by chemical vapor deposition while maintaining excellent magnetic properties. Considering high-quality YIG can only be grown on lattice-matched substrates, 
${\textrm{V[TCNE]}_{2}}$ can be an alternative solution for highly compact integrated magnonic devices. The saturation magnetization of ${\textrm{V[TCNE]}_{2}}$ (90 G) is over an order of magnitude smaller compared with that of YIG (1750 G), leading to the small demagnetizing field and significantly reduced bias magnetic field which is desired for many applications \cite{osborn1945demagnetizing,de2018superconductivity,rogachev2005influence}. On the other hand, the lower material spin density weakens the magnon-photon coupling strength $g$ under the same microwave circuit design and magnon mode volume, making it difficult to reach strong coupling.  

\begin {figure}[htbp]
\centering
\includegraphics[width=\linewidth]{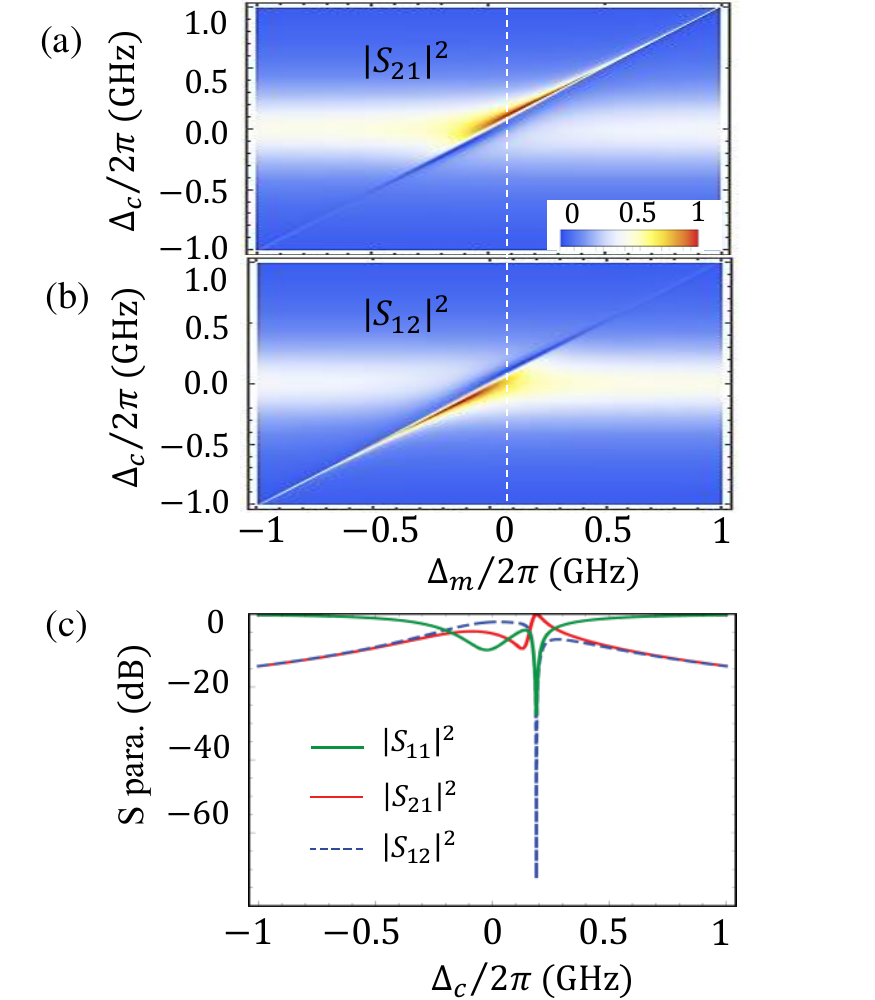}
\caption{(a) \& (b) are the mapping of the transmission ${{|S_{21}|}^{2}}$ and ${{|S_{12}|}^{2}}$ respectively as a function of ${\Delta}_{m}$ and ${\Delta}_{c}$. The ${\textrm{V[TCNE]}_{2}}$ - based circulator is weakly coupled with ${g}/{2\pi}$ = 100 MHz and ${\kappa}_{e}/{2\pi}$ = 600 MHz. Figure (c) plots the full scattering parameters, showing low $|\textrm{IL}|$ and high $|\textrm{iso.}|$ being achieved simultaneously (corresponding the bias condition shown by the dashed white line in (a) \& (b)).}
\label{fig4}
\end {figure}

Here, we discuss the microwave circulation when the system is weakly coupled ($\kappa_{m}<g<\kappa$). For low spin density material, we assign the magnon-photon coupling strength $g/2{\pi}$ to be 100 MHz, with microwave dissipation rates ${{\kappa}_{i}}/2{\pi}=2$ MHz, ${{\kappa}_{e}}/2{\pi}=600$ MHz, and magnon resonant linewidth ${{\kappa}_{m}}/2{\pi}=2$ MHz. The mapping of the transmission scattering parameters is shown in Fig. \ref{fig4}(a) \& (b).  Similar directional transmission between ${|S_{12}|}^{2}$ and ${|S_{21}|}^{2}$ is observed, with a Lorentzian-shaped transparency window that shows the magnon resonance. A line cut of the transmission map at a fixed bias field, indicated by the white dash line in Fig. \ref{fig4}(a) \& (b), is reproduced in \ref{fig4}(c) as a function of the excitation frequency. The low insertion loss (0.09 dB) and high isolation ratio (77 dB) can be achieved with optimized detuning ($\Delta_{m} =$ 0.16 GHz). Due to the small magnon-photon coupling strength, the 20 dB isolation bandwidth is around 0.5 MHz, much narrower compared to that of YIG. This narrow-bandwidth circulator nevertheless has promising applications in circuit QED systems where it can serve as the filtering function for multiplexed superconducting qubits and resonators \cite{ma2014narrowing,krantz2019quantum,heinsoo2018rapid}.

\begin {figure}[htbp]
\centering
\includegraphics[width= 77 mm]{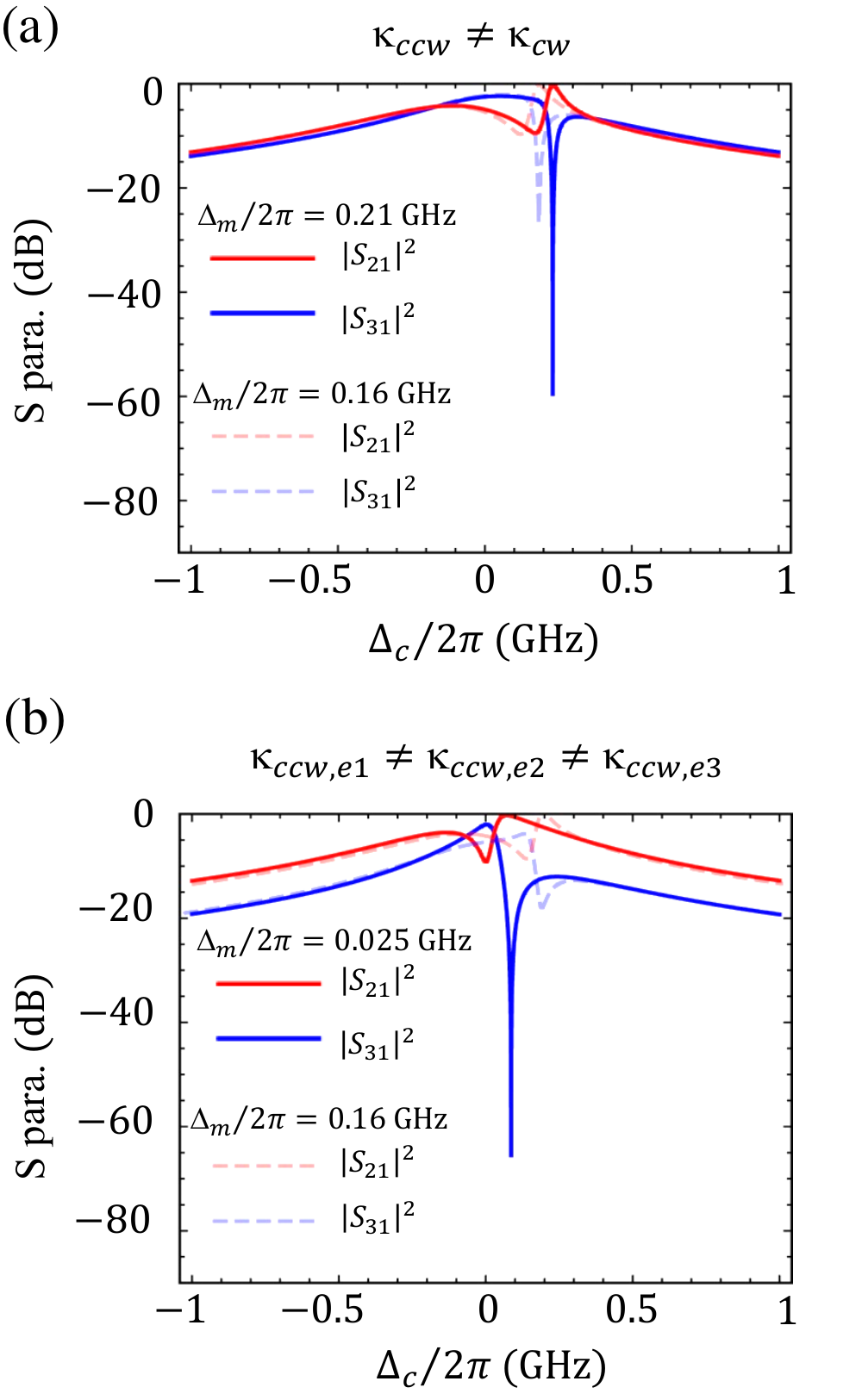}
\caption{(a) Transmission spectra of ${\textrm{V[TCNE]}_{2}}$ - based circulator when the CCW and CW modes have different total external coupling rates (${\kappa_{ccw,e}}/2{\pi}$ $=$ 600 MHz, ${\kappa_{cw,e}}/2{\pi}$ $=$ 720 MHz). The solid lines are transmission S parameters when the $\Delta_{m}$ is 0.21 GHz, while the dashed lines are the transmission parameters when the magnon detuning remains the same as in the symmetrical ${\textrm{V[TCNE]}_{2}}$ circulator. (b). Transmission spectra when the external coupling rates to each waveguides are non-degenerate. The solid and dashed lines denote the S parameters under the optimized and un-optimized magnon detunings, respectively.
}
\label{fig5}
\end {figure}

Next, we discuss the circulation performance for the ${\textrm{V[TCNE]}_{2}}$-based narrow-band circulator, when the three-fold geometrical symmetry cannot be fully satisfied experimentally. With same magnon-photon coupling strength, microwave intrinsic dissipation rates and magnon linewidth, the following two symmetry-breaking conditions are studied. (1). First, the microwave CCW and CW dissipation rates are assigned to be non-degenerate and differ by 20$\%$ (${\kappa_{ccw,e}}/2{\pi}$ $=$ 600 MHz,  ${\kappa_{cw,e}}/2{\pi}$ $=$ 720 MHz). When the system is biased at same magnon detuning ($\Delta_{m} =$ 0.16 GHz) compared with the symmetrical device, as shown in the Fig. \ref{fig5}(a) dashed lines, the isolation ratio decreases to 27 dB. However, by optimizing the magnon detuning to 0.21 GHz, the high isolation (61 dB) and low insertion loss (0.16 dB) can be achieved again at a slightly different frequency, indicating by the solid lines in the Fig. \ref{fig5}(a). (2). Second, the slight differences among external coupling rates between waveguide to the ring resonator can be introduced experimentally, due to the imperfections during device patterning and packaging. Here, the total external coupling rates for CCW and CW modes are designed to be the same (${\kappa_{ccw,e}}$ = ${\kappa_{ccw,e}}$ $=$ ${\kappa_{e}}$ $=$ $2{\pi} \times$ 600 MHz), while the coupling rates to three waveguides are $\frac{1.2}{3}{\kappa_{e}}$, $\frac{1}{3}{\kappa_{e}}$, and $\frac{0.8}{3}{\kappa_{e}}$, respectively. When the magnon detuning $\Delta_{m}$ remains the same with the symmetrical case, illustrated by the dashed line in the Fig. \ref{fig5}(b). The isolation ratio decreases drastically to 19 dB. Similarly, the high performance circulation can be restored via tuning magnon resonances (isolation ratio: 67 dB, insertion loss: 0.15 dB), as shown in the Fig. \ref{fig5}(b) solid lines. It is worth noting that because the system does not maintain the three-fold symmetry, leading to $S_{21} \neq S_{13} \neq S_{32}$, the magnon detuning can only be optimized for one type of connections. Despite of potential system imperfections during the experimental implementations, the tunable magnon resonance can optimize the signal interferences at different ports dynamically to achieve high performance circulation.
%The circulation performance can also be optimized via tuning magnon resonance, thanks to the intrinsic tunability of the magnon excitations.

Also, in the under-coupled scenario, the magnon resonance needs to be tuned close to the microwave resonance, so the overall insertion loss is more sensitive to the magnon linewidth $\kappa_{m}$ than the strongly coupled system. By operating at cryogenic temperatures with reduced ${{\kappa}_{m}}/2{\pi}=0.5$ MHz, the $|\textrm{IL}|$ can be further decreased to $0.04$ dB for the ideal symmetric device and $0.06$ dB for the non-symmetric device as we discussed above, while maintaining $|\textrm{iso.}|>50$ dB.

Recently, a technique for micro-patterning of ${\textrm{V[TCNE]}_{2}}$ thin films has been developed for creating high fidelity lithographically defined structures without observable deterioration of magnetic properties \cite{franson2019low}. With further improvement of the magnon-microwave mode overlap and reduction of magnon mode volume, higher magnon-photon coupling strength can be realized even with low spin density  ${\textrm{V[TCNE]}_{2}}$ system, giving rise to a broader operating bandwidth.

\section{Conclusion}

In summary, we have theoretically investigated a high-performance microwave signal circulation based on a multiport coupled magnon-photon system. The non-reciprocity arises from the interference between the two counter propagating microwave modes, introduced by the chirality-dependent coupling with the magnon excitation. The device implementations are analysed using practical parameters of the superconducting microwave circuit and low Gilbert damping factor materials (YIG \& ${\textrm{V[TCNE]}_{2}}$). High performance microwave circulation with low insertion loss ($< 0.05$ dB) and high isolation ratio ($> 56$ dB) can be achieved with both high/low spin density materials. Although the obtainable isolation bandwidth in low spin density material (${\textrm{V[TCNE]}_{2}}$) is narrower than high spin density platform (YIG), it is beneficial for applications that desire frequency selective isolation such as superconducting quantum computing systems, for example, to isolate the qubits from the environment and to curtail Purcell decay. \cite{ krantz2019quantum,heinsoo2018rapid} Additional advantages of this magnon-photon system include great tunability and directional switchability. The proposed theory is general and can be applied to study multimode induced non-reciprocity in other hybrid systems, such as the quantum optical circulators based on chiral atom-light coupling. \cite{scheucher2016quantum,lodahl2017chiral}.

\section*{Acknowledgements}

We acknowledge funding support from National Science Foundation (EFMA-1741666).  H.X.T. acknowledges support from a Packard Fellowship in Science and Engineering.

\appendix

\section{}

\subsection{Proof of $\boldsymbol{M_2}^{\dagger}\boldsymbol{M_2}=\Gamma_{e}$}
We denote the external and intrinsic dissipation matrices for the microwave signal as $\Gamma_{e}=\left(\begin{array}{ccc}
\kappa_{ccw,e} & 0&0\\
0 & \kappa_{cw,e}&0\\0&0&0
\end{array}\right), $and
$\Gamma_{i}=\left(\begin{array}{ccc}
\kappa_{ccw,i} & 0&0\\
0 & \kappa_{cw,i}&0\\0&0&\kappa_{m}
\end{array}\right). $Consider a special case with no incident wave ($\boldsymbol{s}_{in}=0$) and no intrinsic loss ($\Gamma_{i}=0$). Due to energy conservation, the decay of the intra-cavity energy should equal to the output power $-\frac{d(\boldsymbol{a}^{\dagger}\boldsymbol{a})}{dt}=\boldsymbol{s}_{out}^{\dagger}\boldsymbol{s}_{out}$. From Eq. \ref{eq:refname3}, we have
\begin{equation}
\frac{d(\boldsymbol{a}^{\dagger}\boldsymbol{a})}{dt}=\frac{d\boldsymbol{a}^{\dagger}}{dt}\boldsymbol{a}+\boldsymbol{a}^{\dagger}\frac{d\boldsymbol{a}}{dt}=\boldsymbol{a}^{\dagger}\left(\boldsymbol{M_{1}}^{\dagger}+\boldsymbol{M_{1}}\right)\boldsymbol{a}=-\boldsymbol{a}^{\dagger}\Gamma_{e}\boldsymbol{a}.
\label{eq:refname9}
\end{equation}On the other hand, the output power is 
\begin{equation}
\boldsymbol{s}_{out}^{\dagger}\boldsymbol{s}_{out}=\boldsymbol{a}^{\dagger}\boldsymbol{M_2}^{\dagger}\boldsymbol{M_2}\boldsymbol{a}.
\label{eq:refname10}
\end{equation}
Comparing Eq. \ref{eq:refname9} \& Eq. \ref{eq:refname10} above, we get 
\begin{equation}
\boldsymbol{M_2}^{\dagger}\boldsymbol{M_2}=\Gamma_{e}.
\label{eq:refname11}
\end{equation}

\subsection{Proof of $-\boldsymbol{C}^{\dagger}\boldsymbol{M_2}=\boldsymbol{K}^{\dagger}$}
{}
From energy conservation, we can predict that $\boldsymbol{s}_{in}^{\dagger}\boldsymbol{s}_{in}-\boldsymbol{s}_{out}^{\dagger}\boldsymbol{s}_{out}=\frac{d(\boldsymbol{a}^{\dagger}\boldsymbol{a})}{dt}+\boldsymbol{a}^{\dagger}\Gamma_{i}\boldsymbol{a}.$
Combining Eq. \ref{eq:refname3} Eq. \ref{eq:refname4}  and Eq. \ref{eq:refname11} , we can have 
\begin{equation}
\boldsymbol{s}_{in}^{\dagger}\boldsymbol{C}^{\dagger}\boldsymbol{{M_2}a}+\boldsymbol{a}^{\dagger}\boldsymbol{M_2}^{\dagger}\boldsymbol{C}\boldsymbol{s}_{in}=-\boldsymbol{s}_{in}^{\dagger}\boldsymbol{K}^{\dagger}\boldsymbol{a}-\boldsymbol{a}^{\dagger}\boldsymbol{K}\boldsymbol{s}_{in}.
\label{eq:refname13}
\end{equation}
Therefore, we can get $-\boldsymbol{C}^{\dagger}\boldsymbol{M_2}=\boldsymbol{K}^{\dagger}.$ 

According to the relation among $\boldsymbol{K}$, $\boldsymbol{C}$, and $\boldsymbol{M_2}$,  we can drive the expression of matrix  $\boldsymbol{K}$ and shown in Eq. \ref{eq:refname6}, and the full scattering matrix can be calculated from Eq. \ref{eq:refname8}.

\bibliography{ref}% Produces the bibliography via BibTeX.

%merlin.mbs apsrev4-1.bst 2010-07-25 4.21a (PWD, AO, DPC) hacked
%Control: key (0)
%Control: author (8) initials jnrlst
%Control: editor formatted (1) identically to author
%Control: production of article title (-1) disabled
%Control: page (0) single
%Control: year (1) truncated
%Control: production of eprint (0) enabled
\begin{thebibliography}{71}%
\makeatletter
\providecommand \@ifxundefined [1]{%
 \@ifx{#1\undefined}
}%
\providecommand \@ifnum [1]{%
 \ifnum #1\expandafter \@firstoftwo
 \else \expandafter \@secondoftwo
 \fi
}%
\providecommand \@ifx [1]{%
 \ifx #1\expandafter \@firstoftwo
 \else \expandafter \@secondoftwo
 \fi
}%
\providecommand \natexlab [1]{#1}%
\providecommand \enquote  [1]{``#1''}%
\providecommand \bibnamefont  [1]{#1}%
\providecommand \bibfnamefont [1]{#1}%
\providecommand \citenamefont [1]{#1}%
\providecommand \href@noop [0]{\@secondoftwo}%
\providecommand \href [0]{\begingroup \@sanitize@url \@href}%
\providecommand \@href[1]{\@@startlink{#1}\@@href}%
\providecommand \@@href[1]{\endgroup#1\@@endlink}%
\providecommand \@sanitize@url [0]{\catcode `\\12\catcode `\$12\catcode
  `\&12\catcode `\#12\catcode `\^12\catcode `\_12\catcode `\%12\relax}%
\providecommand \@@startlink[1]{}%
\providecommand \@@endlink[0]{}%
\providecommand \url  [0]{\begingroup\@sanitize@url \@url }%
\providecommand \@url [1]{\endgroup\@href {#1}{\urlprefix }}%
\providecommand \urlprefix  [0]{URL }%
\providecommand \Eprint [0]{\href }%
\providecommand \doibase [0]{http://dx.doi.org/}%
\providecommand \selectlanguage [0]{\@gobble}%
\providecommand \bibinfo  [0]{\@secondoftwo}%
\providecommand \bibfield  [0]{\@secondoftwo}%
\providecommand \translation [1]{[#1]}%
\providecommand \BibitemOpen [0]{}%
\providecommand \bibitemStop [0]{}%
\providecommand \bibitemNoStop [0]{.\EOS\space}%
\providecommand \EOS [0]{\spacefactor3000\relax}%
\providecommand \BibitemShut  [1]{\csname bibitem#1\endcsname}%
\let\auto@bib@innerbib\@empty
%</preamble>
\bibitem [{\citenamefont {Caloz}\ \emph {et~al.}(2018)\citenamefont {Caloz},
  \citenamefont {Al{\`u}}, \citenamefont {Tretyakov}, \citenamefont {Sounas},
  \citenamefont {Achouri},\ and\ \citenamefont
  {Deck-L{\'e}ger}}]{caloz2018electromagnetic}%
  \BibitemOpen
  \bibfield  {author} {\bibinfo {author} {\bibfnamefont {C.}~\bibnamefont
  {Caloz}}, \bibinfo {author} {\bibfnamefont {A.}~\bibnamefont {Al{\`u}}},
  \bibinfo {author} {\bibfnamefont {S.}~\bibnamefont {Tretyakov}}, \bibinfo
  {author} {\bibfnamefont {D.}~\bibnamefont {Sounas}}, \bibinfo {author}
  {\bibfnamefont {K.}~\bibnamefont {Achouri}}, \ and\ \bibinfo {author}
  {\bibfnamefont {Z.-L.}\ \bibnamefont {Deck-L{\'e}ger}},\ }\href@noop {}
  {\bibfield  {journal} {\bibinfo  {journal} {Physical Review Applied}\
  }\textbf {\bibinfo {volume} {10}},\ \bibinfo {pages} {047001} (\bibinfo
  {year} {2018})}\BibitemShut {NoStop}%
\bibitem [{\citenamefont {Linkhart}(2014)}]{linkhart2014microwave}%
  \BibitemOpen
  \bibfield  {author} {\bibinfo {author} {\bibfnamefont {D.~K.}\ \bibnamefont
  {Linkhart}},\ }\href@noop {} {\emph {\bibinfo {title} {Microwave circulator
  design}}}\ (\bibinfo  {publisher} {Artech house},\ \bibinfo {year}
  {2014})\BibitemShut {NoStop}%
\bibitem [{\citenamefont {Fay}\ and\ \citenamefont
  {Comstock}(1965)}]{fay1965operation}%
  \BibitemOpen
  \bibfield  {author} {\bibinfo {author} {\bibfnamefont {C.}~\bibnamefont
  {Fay}}\ and\ \bibinfo {author} {\bibfnamefont {R.}~\bibnamefont {Comstock}},\
  }\href@noop {} {\bibfield  {journal} {\bibinfo  {journal} {IEEE Transactions
  on Microwave Theory and Techniques}\ }\textbf {\bibinfo {volume} {13}},\
  \bibinfo {pages} {15} (\bibinfo {year} {1965})}\BibitemShut {NoStop}%
\bibitem [{\citenamefont {Wu}\ and\ \citenamefont
  {Rosenbaum}(1974)}]{wu1974wide}%
  \BibitemOpen
  \bibfield  {author} {\bibinfo {author} {\bibfnamefont {Y.}~\bibnamefont
  {Wu}}\ and\ \bibinfo {author} {\bibfnamefont {F.~J.}\ \bibnamefont
  {Rosenbaum}},\ }\href@noop {} {\bibfield  {journal} {\bibinfo  {journal}
  {IEEE Transactions on Microwave Theory and Techniques}\ }\textbf {\bibinfo
  {volume} {22}},\ \bibinfo {pages} {849} (\bibinfo {year} {1974})}\BibitemShut
  {NoStop}%
\bibitem [{\citenamefont {Bernier}\ \emph {et~al.}(2017)\citenamefont
  {Bernier}, \citenamefont {Toth}, \citenamefont {Koottandavida}, \citenamefont
  {Ioannou}, \citenamefont {Malz}, \citenamefont {Nunnenkamp}, \citenamefont
  {Feofanov},\ and\ \citenamefont {Kippenberg}}]{bernier2017nonreciprocal}%
  \BibitemOpen
  \bibfield  {author} {\bibinfo {author} {\bibfnamefont {N.~R.}\ \bibnamefont
  {Bernier}}, \bibinfo {author} {\bibfnamefont {L.~D.}\ \bibnamefont {Toth}},
  \bibinfo {author} {\bibfnamefont {A.}~\bibnamefont {Koottandavida}}, \bibinfo
  {author} {\bibfnamefont {M.~A.}\ \bibnamefont {Ioannou}}, \bibinfo {author}
  {\bibfnamefont {D.}~\bibnamefont {Malz}}, \bibinfo {author} {\bibfnamefont
  {A.}~\bibnamefont {Nunnenkamp}}, \bibinfo {author} {\bibfnamefont
  {A.}~\bibnamefont {Feofanov}}, \ and\ \bibinfo {author} {\bibfnamefont
  {T.}~\bibnamefont {Kippenberg}},\ }\href@noop {} {\bibfield  {journal}
  {\bibinfo  {journal} {Nature communications}\ }\textbf {\bibinfo {volume}
  {8}},\ \bibinfo {pages} {604} (\bibinfo {year} {2017})}\BibitemShut {NoStop}%
\bibitem [{\citenamefont {Peterson}\ \emph {et~al.}(2017)\citenamefont
  {Peterson}, \citenamefont {Lecocq}, \citenamefont {Cicak}, \citenamefont
  {Simmonds}, \citenamefont {Aumentado},\ and\ \citenamefont
  {Teufel}}]{peterson2017demonstration}%
  \BibitemOpen
  \bibfield  {author} {\bibinfo {author} {\bibfnamefont {G.~A.}\ \bibnamefont
  {Peterson}}, \bibinfo {author} {\bibfnamefont {F.}~\bibnamefont {Lecocq}},
  \bibinfo {author} {\bibfnamefont {K.}~\bibnamefont {Cicak}}, \bibinfo
  {author} {\bibfnamefont {R.~W.}\ \bibnamefont {Simmonds}}, \bibinfo {author}
  {\bibfnamefont {J.}~\bibnamefont {Aumentado}}, \ and\ \bibinfo {author}
  {\bibfnamefont {J.~D.}\ \bibnamefont {Teufel}},\ }\href@noop {} {\bibfield
  {journal} {\bibinfo  {journal} {Physical Review X}\ }\textbf {\bibinfo
  {volume} {7}},\ \bibinfo {pages} {031001} (\bibinfo {year}
  {2017})}\BibitemShut {NoStop}%
\bibitem [{\citenamefont {Fang}\ \emph {et~al.}(2017)\citenamefont {Fang},
  \citenamefont {Luo}, \citenamefont {Metelmann}, \citenamefont {Matheny},
  \citenamefont {Marquardt}, \citenamefont {Clerk},\ and\ \citenamefont
  {Painter}}]{fang2017generalized}%
  \BibitemOpen
  \bibfield  {author} {\bibinfo {author} {\bibfnamefont {K.}~\bibnamefont
  {Fang}}, \bibinfo {author} {\bibfnamefont {J.}~\bibnamefont {Luo}}, \bibinfo
  {author} {\bibfnamefont {A.}~\bibnamefont {Metelmann}}, \bibinfo {author}
  {\bibfnamefont {M.~H.}\ \bibnamefont {Matheny}}, \bibinfo {author}
  {\bibfnamefont {F.}~\bibnamefont {Marquardt}}, \bibinfo {author}
  {\bibfnamefont {A.~A.}\ \bibnamefont {Clerk}}, \ and\ \bibinfo {author}
  {\bibfnamefont {O.}~\bibnamefont {Painter}},\ }\href@noop {} {\bibfield
  {journal} {\bibinfo  {journal} {Nature Physics}\ }\textbf {\bibinfo {volume}
  {13}},\ \bibinfo {pages} {465} (\bibinfo {year} {2017})}\BibitemShut
  {NoStop}%
\bibitem [{\citenamefont {Metelmann}\ and\ \citenamefont
  {Clerk}(2015)}]{metelmann2015nonreciprocal}%
  \BibitemOpen
  \bibfield  {author} {\bibinfo {author} {\bibfnamefont {A.}~\bibnamefont
  {Metelmann}}\ and\ \bibinfo {author} {\bibfnamefont {A.~A.}\ \bibnamefont
  {Clerk}},\ }\href@noop {} {\bibfield  {journal} {\bibinfo  {journal}
  {Physical Review X}\ }\textbf {\bibinfo {volume} {5}},\ \bibinfo {pages}
  {021025} (\bibinfo {year} {2015})}\BibitemShut {NoStop}%
\bibitem [{\citenamefont {Chapman}\ \emph {et~al.}(2017)\citenamefont
  {Chapman}, \citenamefont {Rosenthal}, \citenamefont {Kerckhoff},
  \citenamefont {Moores}, \citenamefont {Vale}, \citenamefont {Mates},
  \citenamefont {Hilton}, \citenamefont {Lalumiere}, \citenamefont {Blais},\
  and\ \citenamefont {Lehnert}}]{chapman2017widely}%
  \BibitemOpen
  \bibfield  {author} {\bibinfo {author} {\bibfnamefont {B.~J.}\ \bibnamefont
  {Chapman}}, \bibinfo {author} {\bibfnamefont {E.~I.}\ \bibnamefont
  {Rosenthal}}, \bibinfo {author} {\bibfnamefont {J.}~\bibnamefont
  {Kerckhoff}}, \bibinfo {author} {\bibfnamefont {B.~A.}\ \bibnamefont
  {Moores}}, \bibinfo {author} {\bibfnamefont {L.~R.}\ \bibnamefont {Vale}},
  \bibinfo {author} {\bibfnamefont {J.}~\bibnamefont {Mates}}, \bibinfo
  {author} {\bibfnamefont {G.~C.}\ \bibnamefont {Hilton}}, \bibinfo {author}
  {\bibfnamefont {K.}~\bibnamefont {Lalumiere}}, \bibinfo {author}
  {\bibfnamefont {A.}~\bibnamefont {Blais}}, \ and\ \bibinfo {author}
  {\bibfnamefont {K.}~\bibnamefont {Lehnert}},\ }\href@noop {} {\bibfield
  {journal} {\bibinfo  {journal} {Physical Review X}\ }\textbf {\bibinfo
  {volume} {7}},\ \bibinfo {pages} {041043} (\bibinfo {year}
  {2017})}\BibitemShut {NoStop}%
\bibitem [{\citenamefont {Chapman}\ \emph {et~al.}(2019)\citenamefont
  {Chapman}, \citenamefont {Rosenthal},\ and\ \citenamefont
  {Lehnert}}]{chapman2019design}%
  \BibitemOpen
  \bibfield  {author} {\bibinfo {author} {\bibfnamefont {B.~J.}\ \bibnamefont
  {Chapman}}, \bibinfo {author} {\bibfnamefont {E.~I.}\ \bibnamefont
  {Rosenthal}}, \ and\ \bibinfo {author} {\bibfnamefont {K.}~\bibnamefont
  {Lehnert}},\ }\href@noop {} {\bibfield  {journal} {\bibinfo  {journal}
  {Physical Review Applied}\ }\textbf {\bibinfo {volume} {11}},\ \bibinfo
  {pages} {044048} (\bibinfo {year} {2019})}\BibitemShut {NoStop}%
\bibitem [{\citenamefont {Sounas}\ and\ \citenamefont
  {Al{\`u}}(2017)}]{sounas2017non}%
  \BibitemOpen
  \bibfield  {author} {\bibinfo {author} {\bibfnamefont {D.~L.}\ \bibnamefont
  {Sounas}}\ and\ \bibinfo {author} {\bibfnamefont {A.}~\bibnamefont
  {Al{\`u}}},\ }\href@noop {} {\bibfield  {journal} {\bibinfo  {journal}
  {Nature Photonics}\ }\textbf {\bibinfo {volume} {11}},\ \bibinfo {pages}
  {774} (\bibinfo {year} {2017})}\BibitemShut {NoStop}%
\bibitem [{\citenamefont {Stancil}\ and\ \citenamefont
  {Prabhakar}(2009)}]{stancil2009spin}%
  \BibitemOpen
  \bibfield  {author} {\bibinfo {author} {\bibfnamefont {D.~D.}\ \bibnamefont
  {Stancil}}\ and\ \bibinfo {author} {\bibfnamefont {A.}~\bibnamefont
  {Prabhakar}},\ }\href@noop {} {\emph {\bibinfo {title} {Spin waves}}}\
  (\bibinfo  {publisher} {Springer},\ \bibinfo {year} {2009})\BibitemShut
  {NoStop}%
\bibitem [{\citenamefont {Tabuchi}\ \emph {et~al.}(2015)\citenamefont
  {Tabuchi}, \citenamefont {Ishino}, \citenamefont {Noguchi}, \citenamefont
  {Ishikawa}, \citenamefont {Yamazaki}, \citenamefont {Usami},\ and\
  \citenamefont {Nakamura}}]{tabuchi2015coherent}%
  \BibitemOpen
  \bibfield  {author} {\bibinfo {author} {\bibfnamefont {Y.}~\bibnamefont
  {Tabuchi}}, \bibinfo {author} {\bibfnamefont {S.}~\bibnamefont {Ishino}},
  \bibinfo {author} {\bibfnamefont {A.}~\bibnamefont {Noguchi}}, \bibinfo
  {author} {\bibfnamefont {T.}~\bibnamefont {Ishikawa}}, \bibinfo {author}
  {\bibfnamefont {R.}~\bibnamefont {Yamazaki}}, \bibinfo {author}
  {\bibfnamefont {K.}~\bibnamefont {Usami}}, \ and\ \bibinfo {author}
  {\bibfnamefont {Y.}~\bibnamefont {Nakamura}},\ }\href@noop {} {\bibfield
  {journal} {\bibinfo  {journal} {Science}\ }\textbf {\bibinfo {volume}
  {349}},\ \bibinfo {pages} {405} (\bibinfo {year} {2015})}\BibitemShut
  {NoStop}%
\bibitem [{\citenamefont {Zhang}\ \emph {et~al.}(2015)\citenamefont {Zhang},
  \citenamefont {Zou}, \citenamefont {Zhu}, \citenamefont {Marquardt},
  \citenamefont {Jiang},\ and\ \citenamefont {Tang}}]{zhang2015magnon}%
  \BibitemOpen
  \bibfield  {author} {\bibinfo {author} {\bibfnamefont {X.}~\bibnamefont
  {Zhang}}, \bibinfo {author} {\bibfnamefont {C.-L.}\ \bibnamefont {Zou}},
  \bibinfo {author} {\bibfnamefont {N.}~\bibnamefont {Zhu}}, \bibinfo {author}
  {\bibfnamefont {F.}~\bibnamefont {Marquardt}}, \bibinfo {author}
  {\bibfnamefont {L.}~\bibnamefont {Jiang}}, \ and\ \bibinfo {author}
  {\bibfnamefont {H.~X.}\ \bibnamefont {Tang}},\ }\href@noop {} {\bibfield
  {journal} {\bibinfo  {journal} {Nature communications}\ }\textbf {\bibinfo
  {volume} {6}},\ \bibinfo {pages} {8914} (\bibinfo {year} {2015})}\BibitemShut
  {NoStop}%
\bibitem [{\citenamefont {Zhang}\ \emph
  {et~al.}(2016{\natexlab{a}})\citenamefont {Zhang}, \citenamefont {Zou},
  \citenamefont {Jiang},\ and\ \citenamefont {Tang}}]{zhang2016cavity}%
  \BibitemOpen
  \bibfield  {author} {\bibinfo {author} {\bibfnamefont {X.}~\bibnamefont
  {Zhang}}, \bibinfo {author} {\bibfnamefont {C.-L.}\ \bibnamefont {Zou}},
  \bibinfo {author} {\bibfnamefont {L.}~\bibnamefont {Jiang}}, \ and\ \bibinfo
  {author} {\bibfnamefont {H.~X.}\ \bibnamefont {Tang}},\ }\href@noop {}
  {\bibfield  {journal} {\bibinfo  {journal} {Science advances}\ }\textbf
  {\bibinfo {volume} {2}},\ \bibinfo {pages} {e1501286} (\bibinfo {year}
  {2016}{\natexlab{a}})}\BibitemShut {NoStop}%
\bibitem [{\citenamefont {Liu}\ \emph {et~al.}(2016)\citenamefont {Liu},
  \citenamefont {Zhang}, \citenamefont {Tang},\ and\ \citenamefont
  {Flatt{\'e}}}]{liu2016optomagnonics}%
  \BibitemOpen
  \bibfield  {author} {\bibinfo {author} {\bibfnamefont {T.}~\bibnamefont
  {Liu}}, \bibinfo {author} {\bibfnamefont {X.}~\bibnamefont {Zhang}}, \bibinfo
  {author} {\bibfnamefont {H.~X.}\ \bibnamefont {Tang}}, \ and\ \bibinfo
  {author} {\bibfnamefont {M.~E.}\ \bibnamefont {Flatt{\'e}}},\ }\href@noop {}
  {\bibfield  {journal} {\bibinfo  {journal} {Physical Review B}\ }\textbf
  {\bibinfo {volume} {94}},\ \bibinfo {pages} {060405} (\bibinfo {year}
  {2016})}\BibitemShut {NoStop}%
\bibitem [{\citenamefont {Osada}\ \emph {et~al.}(2016)\citenamefont {Osada},
  \citenamefont {Hisatomi}, \citenamefont {Noguchi}, \citenamefont {Tabuchi},
  \citenamefont {Yamazaki}, \citenamefont {Usami}, \citenamefont {Sadgrove},
  \citenamefont {Yalla}, \citenamefont {Nomura},\ and\ \citenamefont
  {Nakamura}}]{osada2016cavity}%
  \BibitemOpen
  \bibfield  {author} {\bibinfo {author} {\bibfnamefont {A.}~\bibnamefont
  {Osada}}, \bibinfo {author} {\bibfnamefont {R.}~\bibnamefont {Hisatomi}},
  \bibinfo {author} {\bibfnamefont {A.}~\bibnamefont {Noguchi}}, \bibinfo
  {author} {\bibfnamefont {Y.}~\bibnamefont {Tabuchi}}, \bibinfo {author}
  {\bibfnamefont {R.}~\bibnamefont {Yamazaki}}, \bibinfo {author}
  {\bibfnamefont {K.}~\bibnamefont {Usami}}, \bibinfo {author} {\bibfnamefont
  {M.}~\bibnamefont {Sadgrove}}, \bibinfo {author} {\bibfnamefont
  {R.}~\bibnamefont {Yalla}}, \bibinfo {author} {\bibfnamefont
  {M.}~\bibnamefont {Nomura}}, \ and\ \bibinfo {author} {\bibfnamefont
  {Y.}~\bibnamefont {Nakamura}},\ }\href@noop {} {\bibfield  {journal}
  {\bibinfo  {journal} {Physical review letters}\ }\textbf {\bibinfo {volume}
  {116}},\ \bibinfo {pages} {223601} (\bibinfo {year} {2016})}\BibitemShut
  {NoStop}%
\bibitem [{\citenamefont {Kostylev}\ \emph {et~al.}(2016)\citenamefont
  {Kostylev}, \citenamefont {Goryach~ev},\ and\ \citenamefont
  {Tobar}}]{kostylev2016superstrong}%
  \BibitemOpen
  \bibfield  {author} {\bibinfo {author} {\bibfnamefont {N.}~\bibnamefont
  {Kostylev}}, \bibinfo {author} {\bibfnamefont {M.}~\bibnamefont
  {Goryach~ev}}, \ and\ \bibinfo {author} {\bibfnamefont {M.~E.}\ \bibnamefont
  {Tobar}},\ }\href@noop {} {\bibfield  {journal} {\bibinfo  {journal} {Applied
  Physics Letters}\ }\textbf {\bibinfo {volume} {108}},\ \bibinfo {pages}
  {062402} (\bibinfo {year} {2016})}\BibitemShut {NoStop}%
\bibitem [{\citenamefont {Bhoi}\ \emph {et~al.}(2014)\citenamefont {Bhoi},
  \citenamefont {Cliff}, \citenamefont {Maksymov}, \citenamefont {Kostylev},
  \citenamefont {Aiyar}, \citenamefont {Venkataramani}, \citenamefont
  {Prasad},\ and\ \citenamefont {Stamps}}]{bhoi2014study}%
  \BibitemOpen
  \bibfield  {author} {\bibinfo {author} {\bibfnamefont {B.}~\bibnamefont
  {Bhoi}}, \bibinfo {author} {\bibfnamefont {T.}~\bibnamefont {Cliff}},
  \bibinfo {author} {\bibfnamefont {I.}~\bibnamefont {Maksymov}}, \bibinfo
  {author} {\bibfnamefont {M.}~\bibnamefont {Kostylev}}, \bibinfo {author}
  {\bibfnamefont {R.}~\bibnamefont {Aiyar}}, \bibinfo {author} {\bibfnamefont
  {N.}~\bibnamefont {Venkataramani}}, \bibinfo {author} {\bibfnamefont
  {S.}~\bibnamefont {Prasad}}, \ and\ \bibinfo {author} {\bibfnamefont
  {R.}~\bibnamefont {Stamps}},\ }\href@noop {} {\bibfield  {journal} {\bibinfo
  {journal} {Journal of Applied Physics}\ }\textbf {\bibinfo {volume} {116}},\
  \bibinfo {pages} {243906} (\bibinfo {year} {2014})}\BibitemShut {NoStop}%
\bibitem [{\citenamefont {Bhoi}\ \emph {et~al.}(2017)\citenamefont {Bhoi},
  \citenamefont {Kim}, \citenamefont {Kim}, \citenamefont {Cho},\ and\
  \citenamefont {Kim}}]{bhoi2017robust}%
  \BibitemOpen
  \bibfield  {author} {\bibinfo {author} {\bibfnamefont {B.}~\bibnamefont
  {Bhoi}}, \bibinfo {author} {\bibfnamefont {B.}~\bibnamefont {Kim}}, \bibinfo
  {author} {\bibfnamefont {J.}~\bibnamefont {Kim}}, \bibinfo {author}
  {\bibfnamefont {Y.-J.}\ \bibnamefont {Cho}}, \ and\ \bibinfo {author}
  {\bibfnamefont {S.-K.}\ \bibnamefont {Kim}},\ }\href@noop {} {\bibfield
  {journal} {\bibinfo  {journal} {Scientific reports}\ }\textbf {\bibinfo
  {volume} {7}},\ \bibinfo {pages} {11930} (\bibinfo {year}
  {2017})}\BibitemShut {NoStop}%
\bibitem [{\citenamefont {Match}\ \emph {et~al.}(2019)\citenamefont {Match},
  \citenamefont {Harder}, \citenamefont {Bai}, \citenamefont {Hyde},\ and\
  \citenamefont {Hu}}]{match2019transient}%
  \BibitemOpen
  \bibfield  {author} {\bibinfo {author} {\bibfnamefont {C.}~\bibnamefont
  {Match}}, \bibinfo {author} {\bibfnamefont {M.}~\bibnamefont {Harder}},
  \bibinfo {author} {\bibfnamefont {L.}~\bibnamefont {Bai}}, \bibinfo {author}
  {\bibfnamefont {P.}~\bibnamefont {Hyde}}, \ and\ \bibinfo {author}
  {\bibfnamefont {C.-M.}\ \bibnamefont {Hu}},\ }\href@noop {} {\bibfield
  {journal} {\bibinfo  {journal} {Physical Review B}\ }\textbf {\bibinfo
  {volume} {99}},\ \bibinfo {pages} {134445} (\bibinfo {year}
  {2019})}\BibitemShut {NoStop}%
\bibitem [{\citenamefont {Harder}\ \emph {et~al.}(2017)\citenamefont {Harder},
  \citenamefont {Bai}, \citenamefont {Hyde},\ and\ \citenamefont
  {Hu}}]{harder2017topological}%
  \BibitemOpen
  \bibfield  {author} {\bibinfo {author} {\bibfnamefont {M.}~\bibnamefont
  {Harder}}, \bibinfo {author} {\bibfnamefont {L.}~\bibnamefont {Bai}},
  \bibinfo {author} {\bibfnamefont {P.}~\bibnamefont {Hyde}}, \ and\ \bibinfo
  {author} {\bibfnamefont {C.-M.}\ \bibnamefont {Hu}},\ }\href@noop {}
  {\bibfield  {journal} {\bibinfo  {journal} {Physical Review B}\ }\textbf
  {\bibinfo {volume} {95}},\ \bibinfo {pages} {214411} (\bibinfo {year}
  {2017})}\BibitemShut {NoStop}%
\bibitem [{\citenamefont {Zhang}\ \emph
  {et~al.}(2014{\natexlab{a}})\citenamefont {Zhang}, \citenamefont {Liu},
  \citenamefont {Flatt{\'e}},\ and\ \citenamefont {Tang}}]{zhang2014electric}%
  \BibitemOpen
  \bibfield  {author} {\bibinfo {author} {\bibfnamefont {X.}~\bibnamefont
  {Zhang}}, \bibinfo {author} {\bibfnamefont {T.}~\bibnamefont {Liu}}, \bibinfo
  {author} {\bibfnamefont {M.~E.}\ \bibnamefont {Flatt{\'e}}}, \ and\ \bibinfo
  {author} {\bibfnamefont {H.~X.}\ \bibnamefont {Tang}},\ }\href@noop {}
  {\bibfield  {journal} {\bibinfo  {journal} {Physical review letters}\
  }\textbf {\bibinfo {volume} {113}},\ \bibinfo {pages} {037202} (\bibinfo
  {year} {2014}{\natexlab{a}})}\BibitemShut {NoStop}%
\bibitem [{\citenamefont {Li}\ \emph {et~al.}(2019)\citenamefont {Li},
  \citenamefont {Polakovic}, \citenamefont {Wang}, \citenamefont {Xu},
  \citenamefont {Lendinez}, \citenamefont {Zhang}, \citenamefont {Ding},
  \citenamefont {Khaire}, \citenamefont {Saglam}, \citenamefont {Divan} \emph
  {et~al.}}]{li2019strong}%
  \BibitemOpen
  \bibfield  {author} {\bibinfo {author} {\bibfnamefont {Y.}~\bibnamefont
  {Li}}, \bibinfo {author} {\bibfnamefont {T.}~\bibnamefont {Polakovic}},
  \bibinfo {author} {\bibfnamefont {Y.-L.}\ \bibnamefont {Wang}}, \bibinfo
  {author} {\bibfnamefont {J.}~\bibnamefont {Xu}}, \bibinfo {author}
  {\bibfnamefont {S.}~\bibnamefont {Lendinez}}, \bibinfo {author}
  {\bibfnamefont {Z.}~\bibnamefont {Zhang}}, \bibinfo {author} {\bibfnamefont
  {J.}~\bibnamefont {Ding}}, \bibinfo {author} {\bibfnamefont {T.}~\bibnamefont
  {Khaire}}, \bibinfo {author} {\bibfnamefont {H.}~\bibnamefont {Saglam}},
  \bibinfo {author} {\bibfnamefont {R.}~\bibnamefont {Divan}},  \emph
  {et~al.},\ }\href@noop {} {\bibfield  {journal} {\bibinfo  {journal}
  {Physical Review Letters}\ }\textbf {\bibinfo {volume} {123}},\ \bibinfo
  {pages} {107701} (\bibinfo {year} {2019})}\BibitemShut {NoStop}%
\bibitem [{\citenamefont {Wang}\ \emph {et~al.}(2019)\citenamefont {Wang},
  \citenamefont {Rao}, \citenamefont {Yang}, \citenamefont {Xu}, \citenamefont
  {Gui}, \citenamefont {Yao}, \citenamefont {You},\ and\ \citenamefont
  {Hu}}]{wang2019nonreciprocity}%
  \BibitemOpen
  \bibfield  {author} {\bibinfo {author} {\bibfnamefont {Y.-P.}\ \bibnamefont
  {Wang}}, \bibinfo {author} {\bibfnamefont {J.}~\bibnamefont {Rao}}, \bibinfo
  {author} {\bibfnamefont {Y.}~\bibnamefont {Yang}}, \bibinfo {author}
  {\bibfnamefont {P.-C.}\ \bibnamefont {Xu}}, \bibinfo {author} {\bibfnamefont
  {Y.}~\bibnamefont {Gui}}, \bibinfo {author} {\bibfnamefont {B.}~\bibnamefont
  {Yao}}, \bibinfo {author} {\bibfnamefont {J.}~\bibnamefont {You}}, \ and\
  \bibinfo {author} {\bibfnamefont {C.-M.}\ \bibnamefont {Hu}},\ }\href@noop {}
  {\bibfield  {journal} {\bibinfo  {journal} {Physical review letters}\
  }\textbf {\bibinfo {volume} {123}},\ \bibinfo {pages} {127202} (\bibinfo
  {year} {2019})}\BibitemShut {NoStop}%
\bibitem [{\citenamefont {Zhang}\ \emph {et~al.}(2019)\citenamefont {Zhang},
  \citenamefont {Galda}, \citenamefont {Han}, \citenamefont {Jin},\ and\
  \citenamefont {Vinokur}}]{zhang2019strong}%
  \BibitemOpen
  \bibfield  {author} {\bibinfo {author} {\bibfnamefont {X.}~\bibnamefont
  {Zhang}}, \bibinfo {author} {\bibfnamefont {A.}~\bibnamefont {Galda}},
  \bibinfo {author} {\bibfnamefont {X.}~\bibnamefont {Han}}, \bibinfo {author}
  {\bibfnamefont {D.}~\bibnamefont {Jin}}, \ and\ \bibinfo {author}
  {\bibfnamefont {V.}~\bibnamefont {Vinokur}},\ }\href@noop {} {\bibfield
  {journal} {\bibinfo  {journal} {arXiv preprint arXiv:1910.14117}\ } (\bibinfo
  {year} {2019})}\BibitemShut {NoStop}%
\bibitem [{\citenamefont {C{\'o}rcoles}\ \emph {et~al.}(2015)\citenamefont
  {C{\'o}rcoles}, \citenamefont {Magesan}, \citenamefont {Srinivasan},
  \citenamefont {Cross}, \citenamefont {Steffen}, \citenamefont {Gambetta},\
  and\ \citenamefont {Chow}}]{corcoles2015demonstration}%
  \BibitemOpen
  \bibfield  {author} {\bibinfo {author} {\bibfnamefont {A.~D.}\ \bibnamefont
  {C{\'o}rcoles}}, \bibinfo {author} {\bibfnamefont {E.}~\bibnamefont
  {Magesan}}, \bibinfo {author} {\bibfnamefont {S.~J.}\ \bibnamefont
  {Srinivasan}}, \bibinfo {author} {\bibfnamefont {A.~W.}\ \bibnamefont
  {Cross}}, \bibinfo {author} {\bibfnamefont {M.}~\bibnamefont {Steffen}},
  \bibinfo {author} {\bibfnamefont {J.~M.}\ \bibnamefont {Gambetta}}, \ and\
  \bibinfo {author} {\bibfnamefont {J.~M.}\ \bibnamefont {Chow}},\ }\href@noop
  {} {\bibfield  {journal} {\bibinfo  {journal} {Nature communications}\
  }\textbf {\bibinfo {volume} {6}},\ \bibinfo {pages} {6979} (\bibinfo {year}
  {2015})}\BibitemShut {NoStop}%
\bibitem [{\citenamefont {Kono}\ \emph {et~al.}(2018)\citenamefont {Kono},
  \citenamefont {Koshino}, \citenamefont {Tabuchi}, \citenamefont {Noguchi},\
  and\ \citenamefont {Nakamura}}]{kono2018quantum}%
  \BibitemOpen
  \bibfield  {author} {\bibinfo {author} {\bibfnamefont {S.}~\bibnamefont
  {Kono}}, \bibinfo {author} {\bibfnamefont {K.}~\bibnamefont {Koshino}},
  \bibinfo {author} {\bibfnamefont {Y.}~\bibnamefont {Tabuchi}}, \bibinfo
  {author} {\bibfnamefont {A.}~\bibnamefont {Noguchi}}, \ and\ \bibinfo
  {author} {\bibfnamefont {Y.}~\bibnamefont {Nakamura}},\ }\href@noop {}
  {\bibfield  {journal} {\bibinfo  {journal} {Nature Physics}\ }\textbf
  {\bibinfo {volume} {14}},\ \bibinfo {pages} {546} (\bibinfo {year}
  {2018})}\BibitemShut {NoStop}%
\bibitem [{\citenamefont {Sparks}(1964)}]{sparks1964ferromagnetic}%
  \BibitemOpen
  \bibfield  {author} {\bibinfo {author} {\bibfnamefont {M.}~\bibnamefont
  {Sparks}},\ }\href@noop {} {\emph {\bibinfo {title} {Ferromagnetic-relaxation
  theory}}}\ (\bibinfo  {publisher} {McGraw-Hill},\ \bibinfo {year}
  {1964})\BibitemShut {NoStop}%
\bibitem [{\citenamefont {Zhang}\ \emph
  {et~al.}(2016{\natexlab{b}})\citenamefont {Zhang}, \citenamefont {Feng},
  \citenamefont {Zhu}, \citenamefont {Yang},\ and\ \citenamefont
  {Li}}]{zhang2016x}%
  \BibitemOpen
  \bibfield  {author} {\bibinfo {author} {\bibfnamefont {Y.}~\bibnamefont
  {Zhang}}, \bibinfo {author} {\bibfnamefont {X.}~\bibnamefont {Feng}},
  \bibinfo {author} {\bibfnamefont {K.}~\bibnamefont {Zhu}}, \bibinfo {author}
  {\bibfnamefont {X.}~\bibnamefont {Yang}}, \ and\ \bibinfo {author}
  {\bibfnamefont {H.}~\bibnamefont {Li}},\ }in\ \href@noop {} {\emph {\bibinfo
  {booktitle} {2016 IEEE International Conference on Microwave and Millimeter
  Wave Technology (ICMMT)}}},\ Vol.~\bibinfo {volume} {1}\ (\bibinfo
  {organization} {IEEE},\ \bibinfo {year} {2016})\ pp.\ \bibinfo {pages}
  {425--427}\BibitemShut {NoStop}%
\bibitem [{\citenamefont {Yu}\ \emph {et~al.}(2014)\citenamefont {Yu},
  \citenamefont {Harberts}, \citenamefont {Adur}, \citenamefont {Lu},
  \citenamefont {Hammel}, \citenamefont {Johnston-Halperin},\ and\
  \citenamefont {Epstein}}]{yu2014ultra}%
  \BibitemOpen
  \bibfield  {author} {\bibinfo {author} {\bibfnamefont {H.}~\bibnamefont
  {Yu}}, \bibinfo {author} {\bibfnamefont {M.}~\bibnamefont {Harberts}},
  \bibinfo {author} {\bibfnamefont {R.}~\bibnamefont {Adur}}, \bibinfo {author}
  {\bibfnamefont {Y.}~\bibnamefont {Lu}}, \bibinfo {author} {\bibfnamefont
  {P.~C.}\ \bibnamefont {Hammel}}, \bibinfo {author} {\bibfnamefont
  {E.}~\bibnamefont {Johnston-Halperin}}, \ and\ \bibinfo {author}
  {\bibfnamefont {A.~J.}\ \bibnamefont {Epstein}},\ }\href@noop {} {\bibfield
  {journal} {\bibinfo  {journal} {Applied Physics Letters}\ }\textbf {\bibinfo
  {volume} {105}},\ \bibinfo {pages} {012407} (\bibinfo {year}
  {2014})}\BibitemShut {NoStop}%
\bibitem [{\citenamefont {Walker}(1957)}]{walker1957magnetostatic}%
  \BibitemOpen
  \bibfield  {author} {\bibinfo {author} {\bibfnamefont {L.~R.}\ \bibnamefont
  {Walker}},\ }\href@noop {} {\bibfield  {journal} {\bibinfo  {journal}
  {Physical Review}\ }\textbf {\bibinfo {volume} {105}},\ \bibinfo {pages}
  {390} (\bibinfo {year} {1957})}\BibitemShut {NoStop}%
\bibitem [{\citenamefont {Li}\ \emph {et~al.}(2018)\citenamefont {Li},
  \citenamefont {Zhu},\ and\ \citenamefont {Agarwal}}]{li2018magnon}%
  \BibitemOpen
  \bibfield  {author} {\bibinfo {author} {\bibfnamefont {J.}~\bibnamefont
  {Li}}, \bibinfo {author} {\bibfnamefont {S.-Y.}\ \bibnamefont {Zhu}}, \ and\
  \bibinfo {author} {\bibfnamefont {G.}~\bibnamefont {Agarwal}},\ }\href@noop
  {} {\bibfield  {journal} {\bibinfo  {journal} {Physical review letters}\
  }\textbf {\bibinfo {volume} {121}},\ \bibinfo {pages} {203601} (\bibinfo
  {year} {2018})}\BibitemShut {NoStop}%
\bibitem [{\citenamefont {Zhang}\ \emph
  {et~al.}(2014{\natexlab{b}})\citenamefont {Zhang}, \citenamefont {Zou},
  \citenamefont {Jiang},\ and\ \citenamefont {Tang}}]{zhang2014strongly}%
  \BibitemOpen
  \bibfield  {author} {\bibinfo {author} {\bibfnamefont {X.}~\bibnamefont
  {Zhang}}, \bibinfo {author} {\bibfnamefont {C.-L.}\ \bibnamefont {Zou}},
  \bibinfo {author} {\bibfnamefont {L.}~\bibnamefont {Jiang}}, \ and\ \bibinfo
  {author} {\bibfnamefont {H.~X.}\ \bibnamefont {Tang}},\ }\href@noop {}
  {\bibfield  {journal} {\bibinfo  {journal} {Physical review letters}\
  }\textbf {\bibinfo {volume} {113}},\ \bibinfo {pages} {156401} (\bibinfo
  {year} {2014}{\natexlab{b}})}\BibitemShut {NoStop}%
\bibitem [{\citenamefont {Pozar}(2009)}]{pozar2009microwave}%
  \BibitemOpen
  \bibfield  {author} {\bibinfo {author} {\bibfnamefont {D.~M.}\ \bibnamefont
  {Pozar}},\ }\href@noop {} {\emph {\bibinfo {title} {Microwave engineering}}}\
  (\bibinfo  {publisher} {John Wiley \& Sons},\ \bibinfo {year}
  {2009})\BibitemShut {NoStop}%
\bibitem [{\citenamefont {Adamyan}\ \emph {et~al.}(2016)\citenamefont
  {Adamyan}, \citenamefont {Kubatkin},\ and\ \citenamefont
  {Danilov}}]{adamyan2016tunable}%
  \BibitemOpen
  \bibfield  {author} {\bibinfo {author} {\bibfnamefont {A.}~\bibnamefont
  {Adamyan}}, \bibinfo {author} {\bibfnamefont {S.}~\bibnamefont {Kubatkin}}, \
  and\ \bibinfo {author} {\bibfnamefont {A.}~\bibnamefont {Danilov}},\
  }\href@noop {} {\bibfield  {journal} {\bibinfo  {journal} {Applied Physics
  Letters}\ }\textbf {\bibinfo {volume} {108}},\ \bibinfo {pages} {172601}
  (\bibinfo {year} {2016})}\BibitemShut {NoStop}%
\bibitem [{\citenamefont {CERN}()}]{cernrf}%
  \BibitemOpen
  \bibfield  {author} {\bibinfo {author} {\bibfnamefont {F.~C.}\ \bibnamefont
  {CERN}},\ }\href@noop {} {\enquote {\bibinfo {title} {Rf engineering basic
  concepts: S-parameters},}\ }\BibitemShut {NoStop}%
\bibitem [{\citenamefont {Bothner}\ \emph {et~al.}(2013)\citenamefont
  {Bothner}, \citenamefont {Knufinke}, \citenamefont {Hattermann},
  \citenamefont {W{\"o}lbing}, \citenamefont {Ferdinand}, \citenamefont
  {Weiss}, \citenamefont {Bernon}, \citenamefont {Fort{\'a}gh}, \citenamefont
  {Koelle},\ and\ \citenamefont {Kleiner}}]{bothner2013inductively}%
  \BibitemOpen
  \bibfield  {author} {\bibinfo {author} {\bibfnamefont {D.}~\bibnamefont
  {Bothner}}, \bibinfo {author} {\bibfnamefont {M.}~\bibnamefont {Knufinke}},
  \bibinfo {author} {\bibfnamefont {H.}~\bibnamefont {Hattermann}}, \bibinfo
  {author} {\bibfnamefont {R.}~\bibnamefont {W{\"o}lbing}}, \bibinfo {author}
  {\bibfnamefont {B.}~\bibnamefont {Ferdinand}}, \bibinfo {author}
  {\bibfnamefont {P.}~\bibnamefont {Weiss}}, \bibinfo {author} {\bibfnamefont
  {S.}~\bibnamefont {Bernon}}, \bibinfo {author} {\bibfnamefont
  {J.}~\bibnamefont {Fort{\'a}gh}}, \bibinfo {author} {\bibfnamefont
  {D.}~\bibnamefont {Koelle}}, \ and\ \bibinfo {author} {\bibfnamefont
  {R.}~\bibnamefont {Kleiner}},\ }\href@noop {} {\bibfield  {journal} {\bibinfo
   {journal} {New Journal of Physics}\ }\textbf {\bibinfo {volume} {15}},\
  \bibinfo {pages} {093024} (\bibinfo {year} {2013})}\BibitemShut {NoStop}%
\bibitem [{\citenamefont {Sierra-Garcia}\ and\ \citenamefont
  {Laurin}(1999)}]{sierra1999study}%
  \BibitemOpen
  \bibfield  {author} {\bibinfo {author} {\bibfnamefont {S.}~\bibnamefont
  {Sierra-Garcia}}\ and\ \bibinfo {author} {\bibfnamefont {J.-J.}\ \bibnamefont
  {Laurin}},\ }\href@noop {} {\bibfield  {journal} {\bibinfo  {journal} {IEEE
  Transactions on antennas and propagation}\ }\textbf {\bibinfo {volume}
  {47}},\ \bibinfo {pages} {58} (\bibinfo {year} {1999})}\BibitemShut {NoStop}%
\bibitem [{\citenamefont {Wu}\ \emph {et~al.}(2008)\citenamefont {Wu},
  \citenamefont {Mu}, \citenamefont {Li},\ and\ \citenamefont
  {Jiao}}]{wu2008design}%
  \BibitemOpen
  \bibfield  {author} {\bibinfo {author} {\bibfnamefont {G.-L.}\ \bibnamefont
  {Wu}}, \bibinfo {author} {\bibfnamefont {W.}~\bibnamefont {Mu}}, \bibinfo
  {author} {\bibfnamefont {D.}~\bibnamefont {Li}}, \ and\ \bibinfo {author}
  {\bibfnamefont {Y.-C.}\ \bibnamefont {Jiao}},\ }\href@noop {} {\bibfield
  {journal} {\bibinfo  {journal} {Progress In Electromagnetics Research}\
  }\textbf {\bibinfo {volume} {78}},\ \bibinfo {pages} {17} (\bibinfo {year}
  {2008})}\BibitemShut {NoStop}%
\bibitem [{\citenamefont {Han}\ and\ \citenamefont
  {Kim}(2002)}]{han2002strong}%
  \BibitemOpen
  \bibfield  {author} {\bibinfo {author} {\bibfnamefont {S.-M.}\ \bibnamefont
  {Han}}\ and\ \bibinfo {author} {\bibfnamefont {Y.-S.}\ \bibnamefont {Kim}},\
  }in\ \href@noop {} {\emph {\bibinfo {booktitle} {Proceedings RAWCON 2002.
  2002 IEEE Radio and Wireless Conference (Cat. No. 02EX573)}}}\ (\bibinfo
  {organization} {IEEE},\ \bibinfo {year} {2002})\ pp.\ \bibinfo {pages}
  {79--82}\BibitemShut {NoStop}%
\bibitem [{\citenamefont {Krawczyk}(2011)}]{krawczyk2011microstrip}%
  \BibitemOpen
  \bibfield  {author} {\bibinfo {author} {\bibfnamefont {M.}~\bibnamefont
  {Krawczyk}},\ }\href@noop {} {\  (\bibinfo {year} {2011})}\BibitemShut
  {NoStop}%
\bibitem [{\citenamefont {Besedin}\ and\ \citenamefont
  {Menushenkov}(2018)}]{besedin2018quality}%
  \BibitemOpen
  \bibfield  {author} {\bibinfo {author} {\bibfnamefont {I.}~\bibnamefont
  {Besedin}}\ and\ \bibinfo {author} {\bibfnamefont {A.~P.}\ \bibnamefont
  {Menushenkov}},\ }\href@noop {} {\bibfield  {journal} {\bibinfo  {journal}
  {EPJ Quantum Technology}\ }\textbf {\bibinfo {volume} {5}},\ \bibinfo {pages}
  {1} (\bibinfo {year} {2018})}\BibitemShut {NoStop}%
\bibitem [{\citenamefont {Chang}\ and\ \citenamefont
  {Hsieh}(2004)}]{chang2004microwave}%
  \BibitemOpen
  \bibfield  {author} {\bibinfo {author} {\bibfnamefont {K.}~\bibnamefont
  {Chang}}\ and\ \bibinfo {author} {\bibfnamefont {L.-H.}\ \bibnamefont
  {Hsieh}},\ }\href@noop {} {\emph {\bibinfo {title} {Microwave ring circuits
  and related structures}}},\ Vol.\ \bibinfo {volume} {156}\ (\bibinfo
  {publisher} {John Wiley \& Sons},\ \bibinfo {year} {2004})\BibitemShut
  {NoStop}%
\bibitem [{\citenamefont {Bahl}(1977)}]{bahl1977designer}%
  \BibitemOpen
  \bibfield  {author} {\bibinfo {author} {\bibfnamefont {I.~J.}\ \bibnamefont
  {Bahl}},\ }\href@noop {} {\  (\bibinfo {year} {1977})}\BibitemShut {NoStop}%
\bibitem [{\citenamefont {Peropadre}\ \emph {et~al.}(2013)\citenamefont
  {Peropadre}, \citenamefont {Zueco}, \citenamefont {Wulschner}, \citenamefont
  {Deppe}, \citenamefont {Marx}, \citenamefont {Gross},\ and\ \citenamefont
  {Garc{\'\i}a-Ripoll}}]{peropadre2013tunable}%
  \BibitemOpen
  \bibfield  {author} {\bibinfo {author} {\bibfnamefont {B.}~\bibnamefont
  {Peropadre}}, \bibinfo {author} {\bibfnamefont {D.}~\bibnamefont {Zueco}},
  \bibinfo {author} {\bibfnamefont {F.}~\bibnamefont {Wulschner}}, \bibinfo
  {author} {\bibfnamefont {F.}~\bibnamefont {Deppe}}, \bibinfo {author}
  {\bibfnamefont {A.}~\bibnamefont {Marx}}, \bibinfo {author} {\bibfnamefont
  {R.}~\bibnamefont {Gross}}, \ and\ \bibinfo {author} {\bibfnamefont {J.~J.}\
  \bibnamefont {Garc{\'\i}a-Ripoll}},\ }\href@noop {} {\bibfield  {journal}
  {\bibinfo  {journal} {Physical Review B}\ }\textbf {\bibinfo {volume} {87}},\
  \bibinfo {pages} {134504} (\bibinfo {year} {2013})}\BibitemShut {NoStop}%
\bibitem [{\citenamefont {Xu}\ \emph {et~al.}(2019)\citenamefont {Xu},
  \citenamefont {Han}, \citenamefont {Fu}, \citenamefont {Zou}, \citenamefont
  {Devoret},\ and\ \citenamefont {Tang}}]{xu2019frequency}%
  \BibitemOpen
  \bibfield  {author} {\bibinfo {author} {\bibfnamefont {M.}~\bibnamefont
  {Xu}}, \bibinfo {author} {\bibfnamefont {X.}~\bibnamefont {Han}}, \bibinfo
  {author} {\bibfnamefont {W.}~\bibnamefont {Fu}}, \bibinfo {author}
  {\bibfnamefont {C.-L.}\ \bibnamefont {Zou}}, \bibinfo {author} {\bibfnamefont
  {M.~H.}\ \bibnamefont {Devoret}}, \ and\ \bibinfo {author} {\bibfnamefont
  {H.~X.}\ \bibnamefont {Tang}},\ }\href@noop {} {\bibfield  {journal}
  {\bibinfo  {journal} {Applied Physics Letters}\ }\textbf {\bibinfo {volume}
  {114}},\ \bibinfo {pages} {192601} (\bibinfo {year} {2019})}\BibitemShut
  {NoStop}%
\bibitem [{\citenamefont {Sun}\ \emph {et~al.}(2012)\citenamefont {Sun},
  \citenamefont {Song}, \citenamefont {Chang}, \citenamefont {Kabatek},
  \citenamefont {Jantz}, \citenamefont {Schneider}, \citenamefont {Wu},
  \citenamefont {Schultheiss},\ and\ \citenamefont {Hoffmann}}]{sun2012growth}%
  \BibitemOpen
  \bibfield  {author} {\bibinfo {author} {\bibfnamefont {Y.}~\bibnamefont
  {Sun}}, \bibinfo {author} {\bibfnamefont {Y.-Y.}\ \bibnamefont {Song}},
  \bibinfo {author} {\bibfnamefont {H.}~\bibnamefont {Chang}}, \bibinfo
  {author} {\bibfnamefont {M.}~\bibnamefont {Kabatek}}, \bibinfo {author}
  {\bibfnamefont {M.}~\bibnamefont {Jantz}}, \bibinfo {author} {\bibfnamefont
  {W.}~\bibnamefont {Schneider}}, \bibinfo {author} {\bibfnamefont
  {M.}~\bibnamefont {Wu}}, \bibinfo {author} {\bibfnamefont {H.}~\bibnamefont
  {Schultheiss}}, \ and\ \bibinfo {author} {\bibfnamefont {A.}~\bibnamefont
  {Hoffmann}},\ }\href@noop {} {\bibfield  {journal} {\bibinfo  {journal}
  {Applied Physics Letters}\ }\textbf {\bibinfo {volume} {101}},\ \bibinfo
  {pages} {152405} (\bibinfo {year} {2012})}\BibitemShut {NoStop}%
\bibitem [{\citenamefont {Zhu}\ \emph {et~al.}(2017)\citenamefont {Zhu},
  \citenamefont {Chang}, \citenamefont {Franson}, \citenamefont {Liu},
  \citenamefont {Zhang}, \citenamefont {Johnston-Halperin}, \citenamefont
  {Wu},\ and\ \citenamefont {Tang}}]{zhu2017patterned}%
  \BibitemOpen
  \bibfield  {author} {\bibinfo {author} {\bibfnamefont {N.}~\bibnamefont
  {Zhu}}, \bibinfo {author} {\bibfnamefont {H.}~\bibnamefont {Chang}}, \bibinfo
  {author} {\bibfnamefont {A.}~\bibnamefont {Franson}}, \bibinfo {author}
  {\bibfnamefont {T.}~\bibnamefont {Liu}}, \bibinfo {author} {\bibfnamefont
  {X.}~\bibnamefont {Zhang}}, \bibinfo {author} {\bibfnamefont
  {E.}~\bibnamefont {Johnston-Halperin}}, \bibinfo {author} {\bibfnamefont
  {M.}~\bibnamefont {Wu}}, \ and\ \bibinfo {author} {\bibfnamefont {H.~X.}\
  \bibnamefont {Tang}},\ }\href@noop {} {\bibfield  {journal} {\bibinfo
  {journal} {Applied Physics Letters}\ }\textbf {\bibinfo {volume} {110}},\
  \bibinfo {pages} {252401} (\bibinfo {year} {2017})}\BibitemShut {NoStop}%
\bibitem [{\citenamefont {Chang}\ \emph {et~al.}(2014)\citenamefont {Chang},
  \citenamefont {Li}, \citenamefont {Zhang}, \citenamefont {Liu}, \citenamefont
  {Hoffmann}, \citenamefont {Deng},\ and\ \citenamefont
  {Wu}}]{chang2014nanometer}%
  \BibitemOpen
  \bibfield  {author} {\bibinfo {author} {\bibfnamefont {H.}~\bibnamefont
  {Chang}}, \bibinfo {author} {\bibfnamefont {P.}~\bibnamefont {Li}}, \bibinfo
  {author} {\bibfnamefont {W.}~\bibnamefont {Zhang}}, \bibinfo {author}
  {\bibfnamefont {T.}~\bibnamefont {Liu}}, \bibinfo {author} {\bibfnamefont
  {A.}~\bibnamefont {Hoffmann}}, \bibinfo {author} {\bibfnamefont
  {L.}~\bibnamefont {Deng}}, \ and\ \bibinfo {author} {\bibfnamefont
  {M.}~\bibnamefont {Wu}},\ }\href@noop {} {\bibfield  {journal} {\bibinfo
  {journal} {IEEE Magnetics Letters}\ }\textbf {\bibinfo {volume} {5}},\
  \bibinfo {pages} {1} (\bibinfo {year} {2014})}\BibitemShut {NoStop}%
\bibitem [{\citenamefont {Onbasli}\ \emph {et~al.}(2014)\citenamefont
  {Onbasli}, \citenamefont {Kehlberger}, \citenamefont {Kim}, \citenamefont
  {Jakob}, \citenamefont {Kl{\"a}ui}, \citenamefont {Chumak}, \citenamefont
  {Hillebrands},\ and\ \citenamefont {Ross}}]{onbasli2014pulsed}%
  \BibitemOpen
  \bibfield  {author} {\bibinfo {author} {\bibfnamefont {M.}~\bibnamefont
  {Onbasli}}, \bibinfo {author} {\bibfnamefont {A.}~\bibnamefont {Kehlberger}},
  \bibinfo {author} {\bibfnamefont {D.}~\bibnamefont {Kim}}, \bibinfo {author}
  {\bibfnamefont {G.}~\bibnamefont {Jakob}}, \bibinfo {author} {\bibfnamefont
  {M.}~\bibnamefont {Kl{\"a}ui}}, \bibinfo {author} {\bibfnamefont
  {A.}~\bibnamefont {Chumak}}, \bibinfo {author} {\bibfnamefont
  {B.}~\bibnamefont {Hillebrands}}, \ and\ \bibinfo {author} {\bibfnamefont
  {C.}~\bibnamefont {Ross}},\ }\href@noop {} {\bibfield  {journal} {\bibinfo
  {journal} {Apl Materials}\ }\textbf {\bibinfo {volume} {2}},\ \bibinfo
  {pages} {106102} (\bibinfo {year} {2014})}\BibitemShut {NoStop}%
\bibitem [{\citenamefont {Dionne}(2009)}]{dionne2009magnetic}%
  \BibitemOpen
  \bibfield  {author} {\bibinfo {author} {\bibfnamefont {G.~F.}\ \bibnamefont
  {Dionne}},\ }\href@noop {} {\emph {\bibinfo {title} {Magnetic oxides}}},\
  Vol.~\bibinfo {volume} {14}\ (\bibinfo  {publisher} {Springer},\ \bibinfo
  {year} {2009})\BibitemShut {NoStop}%
\bibitem [{\citenamefont {Zhu}\ \emph {et~al.}(2016)\citenamefont {Zhu},
  \citenamefont {Zhang}, \citenamefont {Froning}, \citenamefont {Flatt{\'e}},
  \citenamefont {Johnston-Halperin},\ and\ \citenamefont {Tang}}]{zhu2016low}%
  \BibitemOpen
  \bibfield  {author} {\bibinfo {author} {\bibfnamefont {N.}~\bibnamefont
  {Zhu}}, \bibinfo {author} {\bibfnamefont {X.}~\bibnamefont {Zhang}}, \bibinfo
  {author} {\bibfnamefont {I.}~\bibnamefont {Froning}}, \bibinfo {author}
  {\bibfnamefont {M.~E.}\ \bibnamefont {Flatt{\'e}}}, \bibinfo {author}
  {\bibfnamefont {E.}~\bibnamefont {Johnston-Halperin}}, \ and\ \bibinfo
  {author} {\bibfnamefont {H.~X.}\ \bibnamefont {Tang}},\ }\href@noop {}
  {\bibfield  {journal} {\bibinfo  {journal} {Applied Physics Letters}\
  }\textbf {\bibinfo {volume} {109}},\ \bibinfo {pages} {082402} (\bibinfo
  {year} {2016})}\BibitemShut {NoStop}%
\bibitem [{\citenamefont {Franson}\ \emph {et~al.}(2019)\citenamefont
  {Franson}, \citenamefont {Zhu}, \citenamefont {Kurfman}, \citenamefont
  {Chilcote}, \citenamefont {Candido}, \citenamefont {Buchanan}, \citenamefont
  {Flatt{\'e}}, \citenamefont {Tang},\ and\ \citenamefont
  {Johnston-Halperin}}]{franson2019low}%
  \BibitemOpen
  \bibfield  {author} {\bibinfo {author} {\bibfnamefont {A.}~\bibnamefont
  {Franson}}, \bibinfo {author} {\bibfnamefont {N.}~\bibnamefont {Zhu}},
  \bibinfo {author} {\bibfnamefont {S.}~\bibnamefont {Kurfman}}, \bibinfo
  {author} {\bibfnamefont {M.}~\bibnamefont {Chilcote}}, \bibinfo {author}
  {\bibfnamefont {D.~R.}\ \bibnamefont {Candido}}, \bibinfo {author}
  {\bibfnamefont {K.~S.}\ \bibnamefont {Buchanan}}, \bibinfo {author}
  {\bibfnamefont {M.~E.}\ \bibnamefont {Flatt{\'e}}}, \bibinfo {author}
  {\bibfnamefont {H.~X.}\ \bibnamefont {Tang}}, \ and\ \bibinfo {author}
  {\bibfnamefont {E.}~\bibnamefont {Johnston-Halperin}},\ }\href@noop {}
  {\bibfield  {journal} {\bibinfo  {journal} {arXiv preprint arXiv:1910.05325}\
  } (\bibinfo {year} {2019})}\BibitemShut {NoStop}%
\bibitem [{\citenamefont {Wu}(2010)}]{wu2010nonlinear}%
  \BibitemOpen
  \bibfield  {author} {\bibinfo {author} {\bibfnamefont {M.}~\bibnamefont
  {Wu}},\ }in\ \href@noop {} {\emph {\bibinfo {booktitle} {Solid State
  Physics}}},\ Vol.~\bibinfo {volume} {62}\ (\bibinfo  {publisher} {Elsevier},\
  \bibinfo {year} {2010})\ pp.\ \bibinfo {pages} {163--224}\BibitemShut
  {NoStop}%
\bibitem [{\citenamefont {Brandt}(1998)}]{brandt1998superconductor}%
  \BibitemOpen
  \bibfield  {author} {\bibinfo {author} {\bibfnamefont {E.~H.}\ \bibnamefont
  {Brandt}},\ }\href@noop {} {\bibfield  {journal} {\bibinfo  {journal}
  {Physical Review B}\ }\textbf {\bibinfo {volume} {58}},\ \bibinfo {pages}
  {6506} (\bibinfo {year} {1998})}\BibitemShut {NoStop}%
\bibitem [{\citenamefont {De~Gennes}(2018)}]{de2018superconductivity}%
  \BibitemOpen
  \bibfield  {author} {\bibinfo {author} {\bibfnamefont {P.-G.}\ \bibnamefont
  {De~Gennes}},\ }\href@noop {} {\emph {\bibinfo {title} {Superconductivity of
  metals and alloys}}}\ (\bibinfo  {publisher} {CRC Press},\ \bibinfo {year}
  {2018})\BibitemShut {NoStop}%
\bibitem [{\citenamefont {Shapira}\ and\ \citenamefont
  {Neuringer}(1965)}]{shapira1965upper}%
  \BibitemOpen
  \bibfield  {author} {\bibinfo {author} {\bibfnamefont {Y.}~\bibnamefont
  {Shapira}}\ and\ \bibinfo {author} {\bibfnamefont {L.}~\bibnamefont
  {Neuringer}},\ }\href@noop {} {\bibfield  {journal} {\bibinfo  {journal}
  {Physical Review}\ }\textbf {\bibinfo {volume} {140}},\ \bibinfo {pages}
  {A1638} (\bibinfo {year} {1965})}\BibitemShut {NoStop}%
\bibitem [{\citenamefont {Barends}\ \emph {et~al.}(2007)\citenamefont
  {Barends}, \citenamefont {Baselmans}, \citenamefont {Hovenier}, \citenamefont
  {Gao}, \citenamefont {Yates}, \citenamefont {Klapwijk},\ and\ \citenamefont
  {Hoevers}}]{barends2007niobium}%
  \BibitemOpen
  \bibfield  {author} {\bibinfo {author} {\bibfnamefont {R.}~\bibnamefont
  {Barends}}, \bibinfo {author} {\bibfnamefont {J.}~\bibnamefont {Baselmans}},
  \bibinfo {author} {\bibfnamefont {J.}~\bibnamefont {Hovenier}}, \bibinfo
  {author} {\bibfnamefont {J.}~\bibnamefont {Gao}}, \bibinfo {author}
  {\bibfnamefont {S.}~\bibnamefont {Yates}}, \bibinfo {author} {\bibfnamefont
  {T.}~\bibnamefont {Klapwijk}}, \ and\ \bibinfo {author} {\bibfnamefont
  {H.}~\bibnamefont {Hoevers}},\ }\href@noop {} {\bibfield  {journal} {\bibinfo
   {journal} {Ieee Transactions on Applied Superconductivity}\ }\textbf
  {\bibinfo {volume} {17}},\ \bibinfo {pages} {263} (\bibinfo {year}
  {2007})}\BibitemShut {NoStop}%
\bibitem [{\citenamefont {Ebrahimi}\ \emph {et~al.}(2016)\citenamefont
  {Ebrahimi}, \citenamefont {Stallkamp}, \citenamefont {Quint}, \citenamefont
  {Wiesel}, \citenamefont {Vogel}, \citenamefont {Martin},\ and\ \citenamefont
  {Birkl}}]{ebrahimi2016superconducting}%
  \BibitemOpen
  \bibfield  {author} {\bibinfo {author} {\bibfnamefont {M.}~\bibnamefont
  {Ebrahimi}}, \bibinfo {author} {\bibfnamefont {N.}~\bibnamefont {Stallkamp}},
  \bibinfo {author} {\bibfnamefont {W.}~\bibnamefont {Quint}}, \bibinfo
  {author} {\bibfnamefont {M.}~\bibnamefont {Wiesel}}, \bibinfo {author}
  {\bibfnamefont {M.}~\bibnamefont {Vogel}}, \bibinfo {author} {\bibfnamefont
  {A.}~\bibnamefont {Martin}}, \ and\ \bibinfo {author} {\bibfnamefont
  {G.}~\bibnamefont {Birkl}},\ }\href@noop {} {\bibfield  {journal} {\bibinfo
  {journal} {Review of Scientific Instruments}\ }\textbf {\bibinfo {volume}
  {87}},\ \bibinfo {pages} {075110} (\bibinfo {year} {2016})}\BibitemShut
  {NoStop}%
\bibitem [{\citenamefont {Huebl}\ \emph {et~al.}(2013)\citenamefont {Huebl},
  \citenamefont {Zollitsch}, \citenamefont {Lotze}, \citenamefont {Hocke},
  \citenamefont {Greifenstein}, \citenamefont {Marx}, \citenamefont {Gross},\
  and\ \citenamefont {Goennenwein}}]{huebl2013high}%
  \BibitemOpen
  \bibfield  {author} {\bibinfo {author} {\bibfnamefont {H.}~\bibnamefont
  {Huebl}}, \bibinfo {author} {\bibfnamefont {C.~W.}\ \bibnamefont
  {Zollitsch}}, \bibinfo {author} {\bibfnamefont {J.}~\bibnamefont {Lotze}},
  \bibinfo {author} {\bibfnamefont {F.}~\bibnamefont {Hocke}}, \bibinfo
  {author} {\bibfnamefont {M.}~\bibnamefont {Greifenstein}}, \bibinfo {author}
  {\bibfnamefont {A.}~\bibnamefont {Marx}}, \bibinfo {author} {\bibfnamefont
  {R.}~\bibnamefont {Gross}}, \ and\ \bibinfo {author} {\bibfnamefont {S.~T.}\
  \bibnamefont {Goennenwein}},\ }\href@noop {} {\bibfield  {journal} {\bibinfo
  {journal} {Physical Review Letters}\ }\textbf {\bibinfo {volume} {111}},\
  \bibinfo {pages} {127003} (\bibinfo {year} {2013})}\BibitemShut {NoStop}%
\bibitem [{\citenamefont {Maier-Flaig}\ \emph {et~al.}(2017)\citenamefont
  {Maier-Flaig}, \citenamefont {Harder}, \citenamefont {Klingler},
  \citenamefont {Qiu}, \citenamefont {Saitoh}, \citenamefont {Weiler},
  \citenamefont {Gepr{\"a}gs}, \citenamefont {Gross}, \citenamefont
  {Goennenwein},\ and\ \citenamefont {Huebl}}]{maier2017tunable}%
  \BibitemOpen
  \bibfield  {author} {\bibinfo {author} {\bibfnamefont {H.}~\bibnamefont
  {Maier-Flaig}}, \bibinfo {author} {\bibfnamefont {M.}~\bibnamefont {Harder}},
  \bibinfo {author} {\bibfnamefont {S.}~\bibnamefont {Klingler}}, \bibinfo
  {author} {\bibfnamefont {Z.}~\bibnamefont {Qiu}}, \bibinfo {author}
  {\bibfnamefont {E.}~\bibnamefont {Saitoh}}, \bibinfo {author} {\bibfnamefont
  {M.}~\bibnamefont {Weiler}}, \bibinfo {author} {\bibfnamefont
  {S.}~\bibnamefont {Gepr{\"a}gs}}, \bibinfo {author} {\bibfnamefont
  {R.}~\bibnamefont {Gross}}, \bibinfo {author} {\bibfnamefont
  {S.}~\bibnamefont {Goennenwein}}, \ and\ \bibinfo {author} {\bibfnamefont
  {H.}~\bibnamefont {Huebl}},\ }\href@noop {} {\bibfield  {journal} {\bibinfo
  {journal} {Applied Physics Letters}\ }\textbf {\bibinfo {volume} {110}},\
  \bibinfo {pages} {132401} (\bibinfo {year} {2017})}\BibitemShut {NoStop}%
\bibitem [{\citenamefont {Manriquez}\ \emph {et~al.}(1991)\citenamefont
  {Manriquez}, \citenamefont {Yee}, \citenamefont {McLean}, \citenamefont
  {Epstein},\ and\ \citenamefont {Miller}}]{manriquez1991room}%
  \BibitemOpen
  \bibfield  {author} {\bibinfo {author} {\bibfnamefont {J.~M.}\ \bibnamefont
  {Manriquez}}, \bibinfo {author} {\bibfnamefont {G.~T.}\ \bibnamefont {Yee}},
  \bibinfo {author} {\bibfnamefont {R.~S.}\ \bibnamefont {McLean}}, \bibinfo
  {author} {\bibfnamefont {A.~J.}\ \bibnamefont {Epstein}}, \ and\ \bibinfo
  {author} {\bibfnamefont {J.~S.}\ \bibnamefont {Miller}},\ }\href@noop {}
  {\bibfield  {journal} {\bibinfo  {journal} {Science}\ }\textbf {\bibinfo
  {volume} {252}},\ \bibinfo {pages} {1415} (\bibinfo {year}
  {1991})}\BibitemShut {NoStop}%
\bibitem [{\citenamefont {Pokhodnya}\ \emph {et~al.}(2000)\citenamefont
  {Pokhodnya}, \citenamefont {Epstein},\ and\ \citenamefont
  {Miller}}]{pokhodnya2000thin}%
  \BibitemOpen
  \bibfield  {author} {\bibinfo {author} {\bibfnamefont {K.~I.}\ \bibnamefont
  {Pokhodnya}}, \bibinfo {author} {\bibfnamefont {A.~J.}\ \bibnamefont
  {Epstein}}, \ and\ \bibinfo {author} {\bibfnamefont {J.~S.}\ \bibnamefont
  {Miller}},\ }\href@noop {} {\bibfield  {journal} {\bibinfo  {journal}
  {Advanced Materials}\ }\textbf {\bibinfo {volume} {12}},\ \bibinfo {pages}
  {410} (\bibinfo {year} {2000})}\BibitemShut {NoStop}%
\bibitem [{\citenamefont {Osborn}(1945)}]{osborn1945demagnetizing}%
  \BibitemOpen
  \bibfield  {author} {\bibinfo {author} {\bibfnamefont {J.}~\bibnamefont
  {Osborn}},\ }\href@noop {} {\bibfield  {journal} {\bibinfo  {journal}
  {Physical review}\ }\textbf {\bibinfo {volume} {67}},\ \bibinfo {pages} {351}
  (\bibinfo {year} {1945})}\BibitemShut {NoStop}%
\bibitem [{\citenamefont {Rogachev}\ \emph {et~al.}(2005)\citenamefont
  {Rogachev}, \citenamefont {Bollinger},\ and\ \citenamefont
  {Bezryadin}}]{rogachev2005influence}%
  \BibitemOpen
  \bibfield  {author} {\bibinfo {author} {\bibfnamefont {A.}~\bibnamefont
  {Rogachev}}, \bibinfo {author} {\bibfnamefont {A.}~\bibnamefont {Bollinger}},
  \ and\ \bibinfo {author} {\bibfnamefont {A.}~\bibnamefont {Bezryadin}},\
  }\href@noop {} {\bibfield  {journal} {\bibinfo  {journal} {Physical review
  letters}\ }\textbf {\bibinfo {volume} {94}},\ \bibinfo {pages} {017004}
  (\bibinfo {year} {2005})}\BibitemShut {NoStop}%
\bibitem [{\citenamefont {Ma}\ \emph {et~al.}(2014)\citenamefont {Ma},
  \citenamefont {Danilishin}, \citenamefont {Zhao}, \citenamefont {Miao},
  \citenamefont {Korth}, \citenamefont {Chen}, \citenamefont {Ward},\ and\
  \citenamefont {Blair}}]{ma2014narrowing}%
  \BibitemOpen
  \bibfield  {author} {\bibinfo {author} {\bibfnamefont {Y.}~\bibnamefont
  {Ma}}, \bibinfo {author} {\bibfnamefont {S.~L.}\ \bibnamefont {Danilishin}},
  \bibinfo {author} {\bibfnamefont {C.}~\bibnamefont {Zhao}}, \bibinfo {author}
  {\bibfnamefont {H.}~\bibnamefont {Miao}}, \bibinfo {author} {\bibfnamefont
  {W.~Z.}\ \bibnamefont {Korth}}, \bibinfo {author} {\bibfnamefont
  {Y.}~\bibnamefont {Chen}}, \bibinfo {author} {\bibfnamefont {R.~L.}\
  \bibnamefont {Ward}}, \ and\ \bibinfo {author} {\bibfnamefont {D.~G.}\
  \bibnamefont {Blair}},\ }\href@noop {} {\bibfield  {journal} {\bibinfo
  {journal} {Physical review letters}\ }\textbf {\bibinfo {volume} {113}},\
  \bibinfo {pages} {151102} (\bibinfo {year} {2014})}\BibitemShut {NoStop}%
\bibitem [{\citenamefont {Krantz}\ \emph {et~al.}(2019)\citenamefont {Krantz},
  \citenamefont {Kjaergaard}, \citenamefont {Yan}, \citenamefont {Orlando},
  \citenamefont {Gustavsson},\ and\ \citenamefont
  {Oliver}}]{krantz2019quantum}%
  \BibitemOpen
  \bibfield  {author} {\bibinfo {author} {\bibfnamefont {P.}~\bibnamefont
  {Krantz}}, \bibinfo {author} {\bibfnamefont {M.}~\bibnamefont {Kjaergaard}},
  \bibinfo {author} {\bibfnamefont {F.}~\bibnamefont {Yan}}, \bibinfo {author}
  {\bibfnamefont {T.~P.}\ \bibnamefont {Orlando}}, \bibinfo {author}
  {\bibfnamefont {S.}~\bibnamefont {Gustavsson}}, \ and\ \bibinfo {author}
  {\bibfnamefont {W.~D.}\ \bibnamefont {Oliver}},\ }\href@noop {} {\bibfield
  {journal} {\bibinfo  {journal} {Applied Physics Reviews}\ }\textbf {\bibinfo
  {volume} {6}},\ \bibinfo {pages} {021318} (\bibinfo {year}
  {2019})}\BibitemShut {NoStop}%
\bibitem [{\citenamefont {Heinsoo}\ \emph {et~al.}(2018)\citenamefont
  {Heinsoo}, \citenamefont {Andersen}, \citenamefont {Remm}, \citenamefont
  {Krinner}, \citenamefont {Walter}, \citenamefont {Salath{\'e}}, \citenamefont
  {Gasparinetti}, \citenamefont {Besse}, \citenamefont {Poto{\v{c}}nik},
  \citenamefont {Wallraff} \emph {et~al.}}]{heinsoo2018rapid}%
  \BibitemOpen
  \bibfield  {author} {\bibinfo {author} {\bibfnamefont {J.}~\bibnamefont
  {Heinsoo}}, \bibinfo {author} {\bibfnamefont {C.~K.}\ \bibnamefont
  {Andersen}}, \bibinfo {author} {\bibfnamefont {A.}~\bibnamefont {Remm}},
  \bibinfo {author} {\bibfnamefont {S.}~\bibnamefont {Krinner}}, \bibinfo
  {author} {\bibfnamefont {T.}~\bibnamefont {Walter}}, \bibinfo {author}
  {\bibfnamefont {Y.}~\bibnamefont {Salath{\'e}}}, \bibinfo {author}
  {\bibfnamefont {S.}~\bibnamefont {Gasparinetti}}, \bibinfo {author}
  {\bibfnamefont {J.-C.}\ \bibnamefont {Besse}}, \bibinfo {author}
  {\bibfnamefont {A.}~\bibnamefont {Poto{\v{c}}nik}}, \bibinfo {author}
  {\bibfnamefont {A.}~\bibnamefont {Wallraff}},  \emph {et~al.},\ }\href@noop
  {} {\bibfield  {journal} {\bibinfo  {journal} {Physical Review Applied}\
  }\textbf {\bibinfo {volume} {10}},\ \bibinfo {pages} {034040} (\bibinfo
  {year} {2018})}\BibitemShut {NoStop}%
\bibitem [{\citenamefont {Scheucher}\ \emph {et~al.}(2016)\citenamefont
  {Scheucher}, \citenamefont {Hilico}, \citenamefont {Will}, \citenamefont
  {Volz},\ and\ \citenamefont {Rauschenbeutel}}]{scheucher2016quantum}%
  \BibitemOpen
  \bibfield  {author} {\bibinfo {author} {\bibfnamefont {M.}~\bibnamefont
  {Scheucher}}, \bibinfo {author} {\bibfnamefont {A.}~\bibnamefont {Hilico}},
  \bibinfo {author} {\bibfnamefont {E.}~\bibnamefont {Will}}, \bibinfo {author}
  {\bibfnamefont {J.}~\bibnamefont {Volz}}, \ and\ \bibinfo {author}
  {\bibfnamefont {A.}~\bibnamefont {Rauschenbeutel}},\ }\href@noop {}
  {\bibfield  {journal} {\bibinfo  {journal} {Science}\ }\textbf {\bibinfo
  {volume} {354}},\ \bibinfo {pages} {1577} (\bibinfo {year}
  {2016})}\BibitemShut {NoStop}%
\bibitem [{\citenamefont {Lodahl}\ \emph {et~al.}(2017)\citenamefont {Lodahl},
  \citenamefont {Mahmoodian}, \citenamefont {Stobbe}, \citenamefont
  {Rauschenbeutel}, \citenamefont {Schneeweiss}, \citenamefont {Volz},
  \citenamefont {Pichler},\ and\ \citenamefont {Zoller}}]{lodahl2017chiral}%
  \BibitemOpen
  \bibfield  {author} {\bibinfo {author} {\bibfnamefont {P.}~\bibnamefont
  {Lodahl}}, \bibinfo {author} {\bibfnamefont {S.}~\bibnamefont {Mahmoodian}},
  \bibinfo {author} {\bibfnamefont {S.}~\bibnamefont {Stobbe}}, \bibinfo
  {author} {\bibfnamefont {A.}~\bibnamefont {Rauschenbeutel}}, \bibinfo
  {author} {\bibfnamefont {P.}~\bibnamefont {Schneeweiss}}, \bibinfo {author}
  {\bibfnamefont {J.}~\bibnamefont {Volz}}, \bibinfo {author} {\bibfnamefont
  {H.}~\bibnamefont {Pichler}}, \ and\ \bibinfo {author} {\bibfnamefont
  {P.}~\bibnamefont {Zoller}},\ }\href@noop {} {\bibfield  {journal} {\bibinfo
  {journal} {Nature}\ }\textbf {\bibinfo {volume} {541}},\ \bibinfo {pages}
  {473} (\bibinfo {year} {2017})}\BibitemShut {NoStop}%
\end{thebibliography}%

\end{document}